\RequirePackage{fix-cm}
\documentclass[twocolumn]{svjour3}          
\smartqed  
\usepackage{graphicx}
\usepackage{amsmath}
\usepackage{caption}
\usepackage{multirow}
\usepackage{url}
\usepackage{hyperref}
\usepackage{comment}
\usepackage[pdftex,dvipsnames]{xcolor}
\usepackage{xargs}                      

\usepackage[ruled]{algorithm2e}
\usepackage{bm}
\usepackage{amsmath,mathtools}
\usepackage{listings}

\usepackage[colorinlistoftodos,prependcaption,textsize=tiny]{todonotes}
\newcommandx{\unsure}[2][1=]{\todo[linecolor=red,backgroundcolor=red!25,bordercolor=red,#1]{#2}}
\newcommandx{\change}[2][1=]{\todo[linecolor=blue,backgroundcolor=blue!25,bordercolor=blue,#1]{#2}}
\newcommandx{\info}[2][1=]{\todo[linecolor=OliveGreen,backgroundcolor=OliveGreen!25,bordercolor=OliveGreen,#1]{#2}}
\newcommandx{\improvement}[2][1=]{\todo[linecolor=Plum,backgroundcolor=Plum!25,bordercolor=Plum,#1]{#2}}
\newcommandx{\thiswillnotshow}[2][1=]{\todo[disable,#1]{#2}}

\newcommand{\ignore}[1]{}

\definecolor{codegreen}{rgb}{0,0.6,0}
\definecolor{codegray}{rgb}{0.5,0.5,0.5}
\definecolor{codepurple}{rgb}{0.58,0,0.82}
\definecolor{backcolour}{rgb}{0.95,0.95,0.92}

\lstdefinestyle{mystyle}{
    backgroundcolor=\color{backcolour},   
    commentstyle=\color{codegreen},
    keywordstyle=\color{magenta},
    numberstyle=\tiny\color{codegray},
    stringstyle=\color{codepurple},
    basicstyle=\ttfamily\footnotesize,
    breakatwhitespace=false,         
    breaklines=true,                 
    captionpos=b,                    
    keepspaces=true,                 
    numbers=left,                    
    numbersep=5pt,                  
    showspaces=false,                
    showstringspaces=false,
    showtabs=false,                  
    tabsize=2
}

\begin{document}\sloppy

\title{Serverless Streaming for Emerging Media: Towards 5G Network-Driven Cost Optimization \thanks{This work has been realized in the context of the 5G-MEDIA project (\url{www.5gmedia.eu}), which has received funding from the European Union's Horizon 2020 research and innovation programme under grant agreement No. 761699.}
}

\subtitle{A real-time adaptive streaming FaaS service for small-session-oriented immersive media}

\titlerunning{Serverless Streaming for Emerging Media: 5G Network-driven Cost Optimization}        

\author{Konstantinos~Konstantoudakis \and
        David~Breitgand \and
        Alexandros~Doumanoglou \and
        Nikolaos~Zioulis \and
        Avi~Weit \and
        Kyriaki~Christaki \and
        Petros~Drakoulis \and
        Emmanouil~Christakis \and
        Dimitrios~Zarpalas \and
        Petros~Daras}

\authorrunning{K. Konstantoudakis, D. Breitgand et al.} 

\institute{\begin{flushleft}K.~Konstantoudakis, A.~Doumanoglou, N.~Zioulis, K.~Christaki, P.~Drakoulis, E.~Christakis, D.~Zarpalas \\ and P.~Daras  are with: \at
              Visual Computing Lab (VCL),\\
              Information Technologies Institute (ITI), \\
              Centre for Research and Technology - Hellas (CERTH),\\ 
              Thessaloniki, Greece  \\
              Tel.: +30-2310-464160\\
              Fax: +30-2310-464164\\
              \email{k.konstantoudakis@iti.gr; aldoum@iti.gr; nzioulis@iti.gr; kchristaki@iti.gr;\\ petros.drakoulis@iti.gr; manchr@iti.gr; zarpalas@iti.gr; daras@iti.gr}           
           \and
           D.~Breitgand and A.~Weit are with: \at
              Hybrid Cloud, Cloud and Data Technologies, \\
              IBM Research, \\
              Haifa, Israel \\
              Tel.: +972-4-8296211 \\
              \email{davidbr@il.ibm.com; weit@il.ibm.com}\end{flushleft}
}


\maketitle

\begin{abstract}
Immersive 3D media is an emerging type of media that captures, encodes and reconstructs the 3D appearance of people and objects, with applications in tele-presence, teleconference, entertainment, gaming and other fields. In this paper, we discuss a novel concept of live 3D immersive media streaming in a serverless setting. In particular, we present a novel network-centric adaptive streaming framework which deviates from a traditional client-based adaptive streaming used in 2D video. In our framework, the decisions for the production of the transcoding profiles are taken in a centralized manner, by considering consumer metrics vs provisioning costs and inferring an expected consumer quality of experience and behavior based on them. 

\ignore{Furthermore, consumers are directed as to which transcoding profiles to consume by a centralized logic which optimizes consumer QoE while trying to maximize profit for the provider. Costs are being optimized by deploying transcoding units in a serverless 5G infrastructure, with each transcoder being dedicated to produce a single profile. Apart from being among the first works to discuss serverless streaming, another main novelty of this work includes extending a popular open source Function-as-a-Service (FaaS) platform, Apache OpenWhisk, to support media intensive applications. In particular, the framework is extended to allow usage of specialized hardware, such as GPUs for transcoding efficiency as well as inbound network traffic connectivity among serverless functions. A detailed experimental work indicates the effectiveness of the proposed optimization strategy.} In addition, we demonstrate that a naive application of the serverless paradigm might be sub optimal under some common immersive 3D media scenarios.
\end{abstract}

\section{Introduction}
\label{sec:intro}
Media intensive applications and services become increasingly important. 
The ongoing COVID-19 pandemic has forced people to work, learn, and communicate remotely on an unprecedented scale. With more people in quarantine and isolation, the demand for low latency applications, such as video streaming, online games, and teleconferencing has soared to the point that has prompted some countries to look at ways to curb streaming data to avoid overwhelming the Internet~\cite{cnn-slow-down-media}. It has been suggested by many that in a post-COVID-19 world, as restrictions are gradually lifted, many people might use telecommunication as a new normal mode of working. Several large companies have already announced that this unintended pilot on remote teleworking might become the standard way of how people will work in the 21\textsuperscript{st} century.    

With the emergence of immersive media, this option -- remote teleworking and infotainment -- becomes even more attractive and real, since much better quality of experience will be provided to the users. However, immersive media is likely to further exacerbate the issues related to bandwidth and latency (even in the new generation 5G networks), since all next-generation media types~\cite{wien2019standardization} \textemdash either omni-directional ($360^o$) or multi-view or three-dimensional \textemdash impose bandwidth requirements and latency requirements that vastly surpass those of the traditional media, even when the high-end profiles of the current media are considered (i.e. UHD).

A number of approaches aim at mitigating this issue by optimizing the use of resources for media intensive services.
The media services vary by type. A large and increasingly important family of media services are related to tele-presence and infotainment. These services are characterised by highly dynamic consumer populations. Therefore they require  efficient scaling with instantaneous elasticity to handle irregular workload spikes~\cite{soltanian2018ads}.

However, for the real-time media streaming setting, the resource management should extend beyond scaling. The finer-grained decisions might include selection of bit-rate and transcoding profiles to optimize cost-efficiency from the service provider perspective~\cite{zheng2016online}.
More advanced optimization relies on recent standards, such as MPEG-DASH SAND~\cite{mpeg-dash-sand}, which leverages the knowledge it obtains from the network to collaboratively manage media services in order to optimize the users' quality of experience (QoE)~\cite{mehrabi2017joint}.

The aforementioned approaches evolved for the centralized cloud model and are limited by each cloud provider's infrastructure, functionalities, and billing schemes. With the emergence of the 5G networks, ultra-fast, ultra-reliable, and high bandwidth capable edge becomes an attractive option to media services developers. For the immersive media, 5G is a crucial enabling technology, since its targeted key performance indicators stipulated by the architecture documents are essential to providing good QoE for the users~\cite{5g-media-slice-nem}.

Software Defined Networks (SDN) and Network Function Virtualization (NFV) technologies drive the \emph{network softwarization} transformation. A softwarized network is much more amenable to collaborative application and infrastructure optimization via optimized workload placement, application demand adaptation, and network optimization across cloud and edge, based on elaborate monitoring of the infrastructure and service behavior analytics. The cloud-native transformation that drives innovation in the modern cloud and telco cloud edge opens up a number of new opportunities for fine grain resource optimization.

The finer-grained approaches are able to factor in the information provided by the network into the optimization schemes~\cite{zheng2016online} and are better suited to address the central challenge of developing a network architecture being able to dynamically adapt to fluctuating traffic patterns~\cite{ding20185g}.

Serverless computing was first introduced in the end of 2014 and in the last two or three years it has become an extremely popular cloud native pattern used to build highly granular, yet very cost-efficient, micro-services. Serverless computing is an execution model in which a provider of a serverless computing platform manages servers in the back-end and dynamically allocates server resources to virtualization containers (e.g. Docker containers) to execute customers' workloads. In serverless computing, a developer only focuses on the code while the actual packaging and execution is being taken care of by the serverless framework. Broadly speaking, a serverless application scales to zero in absence of the load and automatically scales out (almost instantaneously) when the load is applied. A customer of the serverless computing framework (e.g. a developer) does not have to worry about auto-scaling. This mechanism is automatically included with the serverless framework. A serverless execution model, where a unit of work is a \emph{function} provided on demand (e.g. in response to some \emph{event}) is called Function-as-a-Service (FaaS). FaaS is a sub-model in a broader serverless paradigm. However, exempting instances where clarity demands a specific term, we will use the terms serverless and FaaS interchangeably in this paper.
An important feature of FaaS is its billing model. FaaS comes much closer to the initial business value promise of the Cloud --- pay as you go --- than any other cloud consumption model. A typical billing scheme for FaaS is based on amount of main memory committed during the execution multiplied by the number of seconds, to the granularity of 100 ms  (and is hence priced per GB $\cdot$ seconds). In order to simplify scheduling and flatten capacity planning cycle, FaaS providers limit maximal lifetime of serverless functions by 10--15 minutes. Because of its importance to scheduling (serverless functions are treated as ``sand" that can always be scheduled among ``boulders", i.e. jobs with generally distributed lifetimes), limited maximal lifetime is not a transient feature of FaaS. Therefore, it is mostly suited for session based, event driven, highly dynamic, but relatively short workloads\footnote{Note that there exist popular serverless frameworks, such as Knative~\cite{knative} that do not have limitations on the maximal life time of a function. In these framework, function is not even a building block. Rather such frameworks help building Web services that scale to zero helping with code to container devops cycle along the way.}.

These are exactly the characteristics of many immersive media applications. However, to the best of our knowledge, FaaS is not being widely applied to media intensive services yet. There are multiple reasons for that: First, serverless functions do not communicate with each other via the data network. A typical inter-function communication is via a database. This is way too slow and inadequate for media. Second, FaaS frameworks do not support Day 1, Day 2 configuration of services based on FaaS. Consider an application, in which serverless functions should be executed in response to events, get configured to connect to the rest of an application and then terminate while the rest of an application continues to execute. Such complex management flows are not supported in current frameworks. Third, in many media intensive applications, the use of specialized hardware (e.g. GPUs) is required. This is not supported out of the box neither by open source FaaS, nor by commercial offerings. Fourth, since this model is relatively new, it is largely unknown to the broader community involved with immersive media.

This paper is intended to fill this void. We present a novel architectural approach to developing cost-efficient immersive media applications using the FaaS approach. The overall architectural framework and standards for deploying applications in 5G edge is being evolved by standardization organizations, such as ETSI that stipulates application of FaaS technology in 5G MEC~\cite{etsi-mec}. A typical use case envisioned for FaaS in 5G MEC is IoT. In our previous work, we applied the 5G MEC principles to media intensive applications at the cloud edge~\cite{AlvarezTBC-2019} and presented an overall architectural framework that pioneered the use of FaaS in media intensive applications. That work applied for the first time FaaS to NFV orchestration utilizing a FaaS VIM integrated with the ETSI MANO framework.

In this paper, we continue this line of work by considering session-based workloads typical in immersive media streaming related to infotainment and tele-presence. We developed a fully functional prototype of a tele-immersive gaming service, where time-varying multi-view textured meshes of two players are being produced in real time (a watertight geometry of a player is being produced from four camera streams) and embedded into the virtual environment, where the players can freely move in all 6 degrees of freedom. 
The players communicate with each other via a broker that is being placed in the 5G MEC in geographical proximity to the players to leverage the 5G latency and bandwidth for the sake of the application. Spectators can join from any edge location and also from non 5G access network. The spectators tolerate some small lag (much like it is the case for the sport events broadcasting). 

In contrast to the players, who directly exchange immersive media frames via the broker, the spectators consume 3D streams that are being transcoded to match the capabilities of the spectators' terminals. It should be noted that we take an approach different from a typical media streaming architecture. Rather than letting spectators ask for specific transcoding, our application automatically considers the capabilities of the users' terminals and the network conditions and allocates the most cost-efficient transcoding scheme, trying to balance the trade-off between the cost of transcoders, revenue produced by the spectators and the total benefit for spectators in the form of QoE that motivates them to stay longer in the sessions. In other words, our application optimization strives at achieving maximum profit while providing maximum QoE to spectators. Each player's stream may be transcoded to lower bit-rate versions, namely transcoding profiles, that may be consumed by a multitude of spectators. Cheaper transcoding profiles are being accommodated on CPU, with less RAM and perhaps lower quality configuration, while more expensive ones utilize GPUs (using our extended FaaS framework based on Apache Open-Whisk and Kubernetes).

When in-application events of interest occur (e.g. scoring in an immersive game), a replay serverless function can be executed on demand. The function uses some buffered media to produce a replay clip on and stores it in a low cost cloud storage from which spectators can retrieve it at any time. The number of events happening during the session serves as a proxy to estimate the session popularity with the spectators.

For remote spectators joining at edges where no broker is present, a broker is being started on demand, connected to the main broker, which is being used by the players and each of the transcoded 3D streams is being transmitted by only once to the remote broker, to reduce overall traffic load on the network.

A few important points should be noted about our approach. First, each serverless function in our application has one well defined functionality and a single configuration profile. This greatly simplifies design and operation. Second, thanks to the inbuilt auto-scaling, the application is elastic by design. Third, FaaS is an excellent match for the session-based nature of the application and its fine granularity (a single function level) allows to optimize cost-efficiency of resource allocation at the level of individual sessions paying only for what is actually being used. These advantages are not available out of the box in any other cloud-native model.

We validate our approach via extensive experimentation, contrasting our network-centric optimization approach with a naive serverless implementation (which would always start transcoders on demand irrespective of the predicted accrued benefit), and a traditional Virtual Machine (VM) based approach. Since some features (e.g. support for GPUs) are not yet available in commercial offerings, the billing schemes necessary for experimentation on cost efficiency are not available. To that end, we examine how GPUs are being offered in the cloud today and examine conditions for their cost efficiency in FaaS offerings in 5G MEC. We then use the billing schemes extrapolated from this study as a proxy to obtain preliminary figures illustrating a comparative cost-efficiency of the proposed approach.

In summary, our main contributions are as follows:
\begin{itemize}
    \item We expand the range of applications for serverless architectures to media streaming, addressing its requirements and introducing the concept of serverless streaming; 
    \item We apply this concept to a demanding use-case of next-generation media by implementing and deploying an adaptive streaming service to 5G-enabled network infrastructure, in the context of a real-time and interactive media scenario; 
    \item We show how a serverless architecture within a 5G framework can also enable in-network service optimization and network centric adaptation for the media intensive verticals;
    \item We demonstrate the cost effectiveness of serverless streaming compared to traditional solutions taking into account the balance between the total QoE and cost of production;  
    \item Our findings also serve as a guideline to how serverless should be used in similar use-cases and indicate that naively applying serverless would be sub-optimal.
\end{itemize}

The rest of this paper is organized as follows: In Section~\ref{sec:related} we discuss related work and the present work's relation and connection to it. In Section~\ref{extended-faas}, we outline our extensions to Apache OpenWhisk serverless framework, while in Section~\ref{sec:serverless} we present our serverless adaptive streaming service. In Section~\ref{sec:formulation}, network-centric cost optimization is discussed and in Section \ref{sec:experiments} experimental results are given. Finally, Section~\ref{sec:conclusion} concludes the paper.

\section{Related Work}
\label{sec:related}
In this work, we expand on the novel concept of network-centric 3D immersive media real-time adaptive streaming in a serverless setting. The concept is multidisciplinary and therefore has several partial overlaps with various topics in the literature. 

To facilitate the reader’s comprehension, we split this section into a small number of more focused subsections covering different sub-topics. In Subsection~\ref{sec:immersive_media_platforms}, we briefly describe and provide examples of 3D immersive media production platforms. In Subsection~\ref{sec:adaptive_streaming}, 
we present the principal ideas and some of the more recent advancements in the area of video adaptive streaming. In Subsection~\ref{sec:immersive_media_adaptive_streaming}, we focus on immersive media, namely $360^o$ video and 3D representations. 
Next, in Subsection~\ref{sec:other_adaptive_streaming},  we go over other adaptive streaming solutions, including some of the more recent works in the area of server-based, network-assisted adaptive streaming and cloud-based streaming solutions. Finally, in Subsection~\ref{sec:serverless_overview} we provide an overview of the serverless computing model focusing on the features that are more relevant to the context of this work.

\subsection{3D Immersive Media Production Platforms}
\label{sec:immersive_media_platforms}

The key enabler of 3D immersive media production is a volumetric capturing system. 
A volumetric capturing setup is usually comprised of a $360^o$ arrangement of inward looking camera sensors, defining a capturing space with specific boundaries. 
Despite the fact that volumetric capturing systems most commonly output a multi-view plus depth~\cite{mpeg_i} representation of the captured scene, the most common 3D immersive media format is colored point clouds or textured 3D meshes. 
The latter are produced by 3D reconstruction algorithms~\cite{kazhdan2005reconstruction,kazhdan2006poisson} run on the 3D points of the spatially aligned captured views. 
In general, the 3D reconstruction process can be performed either offline, or in real-time which --- given sufficient computational and network resources --- can additionally allow for live streaming. 

An open, free-to-use, state-of-the-art, low-cost and portable volumetric capture system, which does not integrate a 3D reconstruction algorithm, is~\cite{sterzentsenko_low-cost_2018}. 
One of the earliest low-cost platforms~\cite{zioulis_3d_2016} utilized 4 consumer grade RGB-D sensors and incorporated 3D reconstruction, enabling tele-immersion at interactive rates. 

More recently, Holoportation~\cite{orts-escolano_holoportation_2016} utilized 16 IR-stereo pairs for depth estimation along with 8 color cameras for texturing, to produce and stream high quality 3D textured meshes. 
Even though this system produces stunning 3D reconstructions, its computational complexity is high as it requires 1 GPU per IR stereo pair and a main workstation equipped with 2 GPUs to undertake the task of actual processing.
Moreover, the output bit-stream requires approximately 40 Mbit/frame, which for a 30 frames-per-second real-time streaming scenario would require over 1 Gbps of bandwidth. 

A significant improvement on the volume of the streamable content, which has been kept below 16 Mbps without compromising quality, has been demonstrated by the offline immersive media platform in~\cite{collet_high-quality_2015}. 
To achieve such a remarkable performance, the authors employed 61 12-core Intel Xeon machines, while processing would take 25-29 sec/frame.  Other 3D immersive media platforms also exist in the literature~\cite{dou_fusion4d_2016,schreer_capture_2019}. 
Common elements among most existing works are the increased processing power required to achieve high quality content and the extreme bandwidth requirements for streaming, which can only be mitigated by devoting even more processing resources.

\subsection{Adaptive Streaming}
\label{sec:adaptive_streaming}

Consumers of media content over the internet are highly heterogeneous.
A consumer is characterized by device capabilities, available processing power and network quality (bandwidth, latency, and loss rate). 
The most common way that the contemporary technology optimizes QoE for consumers, is through HTTP Adaptive Streaming (HAS)~\cite{has_survey}.
The objective of HAS is to maintain the viewer’s QoE at high levels, countering the negative impact of the network bandwidth fluctuations. 
In HAS, prior to the distribution, the video needs to be available in segments and encoded in multiple qualities.
The most popular HAS protocols today are MPEG-Dynamic Adaptive Streaming over HTTP (DASH)~\cite{sodagar_mpeg-dash_2011} and Apple’s HTTP Live Streaming (HLS)~\cite{pantos_http_2017}. 
While they have differences in specification and content deployment, recent MPEG’s standardization efforts for the Common Media Application Format (CMAF)~\cite{noauthor_common_nodate}, allow adaptive streaming using either MPEG-DASH or HLS from a single source. 

There exist multiple studies on QoE in video adaptive streaming \cite{seufert_survey_2015,hosfeld_identifying_2015,barman_qoe_2019,paudyal_impact_2016}. Some of the more important factors affecting QoE include: initial delay, stalling frequency, stalling duration, adaptation (quality change) interval, adaptation frequency, adaptation direction, adaptation amplitude, video’s spatial resolution, video’s frame-rate and video’s visual quality~\cite{seufert_survey_2015}. 
Due to the multiplicity of factors affecting a consumer’s QoE, there is no single QoE model that different studies converge on and which can serve as a common reference framework.

HAS leaves encoding schemes and the adaptation strategy without a specification. 
According to~\cite{bentaleb_survey_2019}, and based on the location of the adaptation logic inside the HAS system, HAS schemes can be split into four categories: i) client-based ii) server-based iii) network-assisted and iv) hybrid. 
The most common scheme is (i), in which the adaptation logic runs on the client with the video player fetching the video segments based on a manifest.
In most implementations, the adaptation logic relies on monitoring internal buffer levels and measuring throughput \cite{bhargava_comparative_2019}.
Current state-of-the-art client-based bitrate adaptation algorithms are presented in \cite{spiteri_bola_2016} and \cite{li_probe_2014}. A cutting-edge reinforcement-learning approach is provided by~\cite{mao_neural_2017}, while~\cite{zhang_ensemble_2019} describes ensemble algorithms tailoring different network conditions.

\subsection{Immersive Media Adaptive Streaming}
\label{sec:immersive_media_adaptive_streaming}
\subsubsection{Omnidirectional Media}
\label{sec:omnidirectional_adaptive_streaming}

A survey on $360^o$ video streaming can be found in~\cite{Fan_Lo_Pai_Hsu_2019}.
Regarding immersive media, MPEG has recently standardized the Omnidirectional Media Format (OMAF)~\cite{hannuksela_overview_2019} specification for 360 degree video streaming. 
For the $360^o$ video streaming, the most common adaptation strategy is viewport-based, in which the equirectangular image is split into tiles which are encoded at different bitrates based on the viewing direction of the client~\cite{singla_subjective_2019,schatz_tile-based_2019,schatz_towards_2017,skupin_viewport-dependent_2017,graf_towards_2017} often exploiting tiling support in video coding algorithms, like HEVC~\cite{sullivan_overview_2012,misra_overview_2013}. 

In~\cite{Hosseini_Swaminathan_2016}, a tile-based approach is described using MPEG-DASH SRD (Spatial Relationship Descriptor) and tile-over-viewport prioritization.
Naive tile-based approaches download the portion of the video that the viewer is looking at.
However, the fetching of new tiles from network results in more latency than motion-to-photon latency of the VR headset. 
In~\cite{Xie_Xu_Ban_Zhang_Guo_2017}, a probabilistic approach is taken for a tile pre-fetching strategy that minimizes expected distortion of the downloaded tiles. 
In~\cite{Chakareski_Aksu_Corbillon_Simon_Swaminathan_2018}, a tile-based probabilistic approach is taken, that captures the likelihood of the viewer navigating towards specific tiles in the form of heatmaps. 
In~\cite{He_Qureshi_Qiu_Li_Li_Han_2018}, the authors attempt to provide a solution to $360^o$ video streaming to smartphones, overcoming their processing power limitations compared to desktop PCs. 
Finally,~\cite{Ballard_Griwodz_Steinmetz_Rizk_2019} presents a real-time streaming system of $360^o$ video relying on GPU-based HEVC~\cite{sullivan_overview_2012} coding. 

\subsubsection{3D Media}
\label{sec:3d_media_adaptive_streaming}

Due to a higher complexity of 3D representations, the 3D Immersive media coding and streaming approaches are less mature compared to $360^o$ or standard 2D video.
To begin with, there exist very few 3D immersive media codecs exploiting inter-frame redundancy in time-varying mesh sequences (the mesh sequences of varying geometry and connectivity like the ones produced by real-time 3D reconstruction systems)~\cite{Doumanoglou_Alexiadis_Zarpalas_Daras_2014},~\cite{Yamasaki_Aizawa_2010}. 
Thus, for the 3D mesh geometry, only static 3D mesh codecs are utilized~\cite{Doumanoglou_Drakoulis_Zioulis_Zarpalas_Daras_2019}.
Furthermore, there is very little literature regarding QoE for 3D immersive media streaming, which could drive adaptive streaming systems~\cite{doumanoglou_quality_2018}. 

On the other hand, for the point-cloud representations more options exist. 
In~\cite{Krivokuca_Chou_Koroteev_2020} and~\cite{Chou_Koroteev_Krivokuca_2020}, point clouds are compressed exploiting volumetric function representations, while in~\cite{Mekuria_Blom_Cesar_2017} point cloud sequences are intra-frame and inter-frame coded  based on octrees and motion prediction. 
A detailed survey summarizing works in 3D geometry compression can be found in~\cite{Maglo_Lavoue_Dupont_Hudelot_2015}. 
Finally, the accompanied textures used to colorize the 3D mesh are often compressed using standard 2D image or video compression algorithms like Motion-JPEG (MJPEG) or HEVC.

One of the first works for real-time adaptive streaming of textured 3D time-varying meshes is~\cite{Crowle_Doumanoglou_Poussard_Boniface_Zarpalas_Daras_2015}, which is based on a dynamic rule adaptation strategy modifying compression parameters of the real-time stream. 
For point-cloud streaming, in~\cite{Hosseini_2019} and~\cite{Park_Chou_Hwang_2018}, multiple 3D objects of the same scene are streamed with adaptation relying on content's proximity to the viewer, along with the viewer's looking direction and distance to content. 
One of the first complete HAS implementations of a point-cloud adaptive streaming solution is presented in~\cite{van_der_hooft_towards_2019}. 
The authors in~\cite{van_der_hooft_towards_2019}, demonstrate a DASH-compliant HAS system for dynamic point clouds, introducing rate adaptation heuristics that are based on viewer's position and looking direction, network bandwidth and buffer status. 
At the same time, the encoding scheme utilizes the recently introduced MPEG Video Point Cloud Coding (V-PCC) algorithm~\cite{Cui_Mekuria_Preda_Jang_2019}.

\subsection{Other Adaptive Streaming Solutions}
\label{sec:other_adaptive_streaming}
Server-based, network-assisted and hybrid approaches to adaptive streaming used to be less popular, but recently they started attracting an increased interest with the emergence of Software-Defined-Networking (SDN) and 5G Networks. 
In~\cite{El_Marai_Taleb_Menacer_Koudil_2018}, a DASH-based server-client adaptive streaming system for standard 2D video is proposed that achieves efficiency, stability, fairness and convergence with server and clients co-operating for maximum gains. 
A SAND-DASH network assisted approach is given in~\cite{Mehrabi_Siekkinen_Yla-Jaaski_2019}, describing a method to perform adaptive video streaming to mobile devices, in Multi-Access Edge Computing (MEC) scenarios. Noticeable performance improvements have been observed when the achievable throughput was moderately high or the link qualities across mobile clients were alike.
Finally, in~\cite{Tian_Babcock_Taylor_Ji_2018}, the authors propose Cloud Live Video Streaming (CLVS), 
a model that exploits Amazon S3's storage capabilities in order to enable cost-efficiency in a live video streaming scenario oriented towards small streaming sessions.

The solution in~\cite{Tian_Babcock_Taylor_Ji_2018} eliminates the need for a constantly up-and-running streaming server (and in that sense it is serverless). Rather, a source video is being recorded by a mobile device, on which then it is segmented and encoded. Next, those video segments are pushed into a designated Amazon S3 bucket. On end-user devices, the client program of CLVS directly retrieves the most recent video segments from the S3 bucket, then performs decoding and video playing back. While being inventive and accruing the cost-efficiency advantages compared to a typical solution, in which a cloud based video streaming server can have idle periods, CLVS will not scale to support 3D adaptive streaming neither from the latency, nor from the bandwidth, nor from the cost-efficiency perspectives. Also this design does not allow a network-centric adaptation of QoE. 


In addition to the previously presented related work, we find it important to also mention two other works in the literature that are relevant to our domain.
In~\cite{Kim_Yang_Choi_Lee_Yoon_Kim_Park_2018} a game engine plugin is designed based on MPEG-DASH~\cite{sodagar_mpeg-dash_2011} SRD and HEVC~\cite{sullivan_overview_2012} with an in-game 360 virtual camera in order to enable $360^o$ video streaming of the game environment in e-sports events. 
And in~\cite{Sun_Duanmu_Liu_Wang_Ye_Shi_Dai_2018}, the authors try to exploit the 5G network infrastructure to offer better QoE in $360^o$ video streaming.




\subsection{Serverless Computing for Media}
\label{sec:serverless_overview}
The serverless programming model rapidly becomes popular with developers. Serverless computing relieves the developers from the tasks related to application packaging and server provisioning.  The developers need only to provide the code of their application and a source to executable pipeline automatically creates a running task in a cloud. As explained in the previous section, broadly speaking, the serverless computing paradigm refers to services that scale to zero. FaaS, a specifically popular serverless computing model, refers to structuring applications as stateless functions that are being called on demand (e.g. in response to events). The reader is advised to consult~\cite{KritikosSkzypek-UCC-Companion2018} for a comprehensive review of serverless frameworks.

Recently FaaS has been applied by practitioners to video streaming~\cite{aws-serverless-vod,serverless-video-streaming}. Little scientific literature exists on the topic. In~\cite{Zhang-NOSSDAV19} a measurement study of transcoding tasks has been performed to explore how different lambda function configurations (in terms of memory and proportionally allocated CPU) affect performance and cost.The study reveals that the memory configuration for cost-efficient serverless functions is non-trivial. The best memory configuration is influenced by the task type or even the video content. More work is needed to design an efficient and adaptive system to find the best configuration for serverless functions in video processing pipelines.
In~\cite{Ao_Izhikevich_Voelker_Porter_2018}, a serverless framework facilitating development of video processing pipelines is described. Common to all these solutions is rising serverless functions (e.g. AWS Lambda) for performing operations (e.g. transcoding) on a video file that is uploaded to the cloud storage (e.g. S3 bucket). Upon the file upload, an event is being generated by the storage, which triggers the execution of a serverless function, whose output is either stored in the cloud storage again (potentially creating another trigger for another function execution) or propagated to a Content Distribution Network (CDN), such as AWS CloudFront.

To the best of our knowledge, until this work, no attempt has been made to use serverless functions for adaptive transcoding of a live 3D immersive media stream. In our implementation we used open source Apache OpenWhisk project~\cite{openwhisk}. We leveraged OpenWhisk's capability of executing functions on top of Kubernetes to provide features such as direct network communication among serverless functions (as opposed to the communication via storage or database, which is typical in other frameworks), support for Day 0, Day 1, and Day 2 configuration, as well as support for GPUs. All these features are not being provided out of the box to the developers, which hinders serverless adoption for media intensive applications. In this paper we demonstrate how adding this features might open up new opportunities to achieve cost-efficient immersive media implementations.

\subsection{Summary of our innovations}
\label{sec:summary-of-innovations}
In this paper, we expand on a preliminary design~\cite{doumanoglou2018system} of a novel multiplayer tele-immersive game application~\cite{christaki2019space} where players are embedded inside the game environment via their 3D reconstructed avatars. The gaming application is supported by a 3D immersive media production platform which uses \cite{sterzentsenko_low-cost_2018} for volumetric capturing and a re-implementation of the 3D reconstruction algorithm found in \cite{alexiadis_integrated_2017}. This platform is low-cost, portable, real-time and produces streamable content of $\sim 50$ Mbps,  at interactive rates (25 frames per second).
The application offers live spectating of the game action and on-demand viewing of replay clips. It is deployed on 5G serverless infrastructure and employs adaptive streaming techniques to stream the 3D appearance of the players to spectators. 
Our adaptive streaming algorithm is based on \cite{Doumanoglou_Drakoulis_Zioulis_Zarpalas_Daras_2019} for compressing geometry and MJPEG for compressing textures. Adaptation is achieved by varying compression parameters to produce different profiles at various bit-rates. Further, apart from costs, adaptation optimization is driven by a variant of the QoE model in \cite{zadtootaghaj2018modeling}.
Our work is among the first to discuss live 3D immersive media streaming under a 5G, serverless framework. Such an attempt was not possible before, mainly due serverless frameworks lacking support for network communications of serverless functions, Day 0/1/2 configurations and GPUs. Furthermore, this work is also among the first to provide a network-centric novel adaptive streaming algorithm which takes into account the serverless benefits in order to minimize service costs while offering high QoE to spectators.

\section{FaaS Extensions}

\label{extended-faas}

FaaS frameworks and offerings are rapidly proliferating. However, there are just a few industrial grade open source FaaS platforms available. One such framework is Apache OpenWhisk~\cite{openwhisk}, which powers the IBM Cloud Functions commercial offering~\cite{ibm-cloudfunctions}. Presently, FaaS commercial offerings do not offer usage of GPUs in serverless functions. The reason for that is that GPU sharing is a relatively new topic that poses a number of challenges. Since NVIDIA has introduced Multi-Process Service (MPS) in its Volta GPU architecture~\cite{nvidia-mps}, GPU sharing is being a hot research topic~\cite{DBLP:journals/corr/abs-2005-02088}.

A most common compute virtualization technology powering FaaS is containers. In production, containers are managed by container orchestrators, such as Kubernetes~\cite{kubernetes}. However, current container orchestrators do not know how to leverage architectures such as NVIDIA MPS yet. Thus, a solution that we adopt for extending FaaS to use GPUs is time-sharing of GPUs rather than collocating workloads on the same GPU. Another reason for preferring time sharing to spatial collocation is that collocating workloads on GPUs might require re-writing of the application code. 

Another problem that is currently not being addressed by the FaaS frameworks is supporting both inbound and outbound network traffic to and from serverless functions. Usually, only the outbound traffic is being supported seamlessly. For the inbound traffic, an image of the serverless function container should include some communication service, which might be difficult to do due to inability to expose the function as a service to the outside world and intricate firewall settings. In our solution we rely on using Container Network Interface (CNI) to connect serverless functions to a logical network maintained by container orchestrator. 

\begin{figure*}[!htbp]
    \centering
    \includegraphics[width=\textwidth]{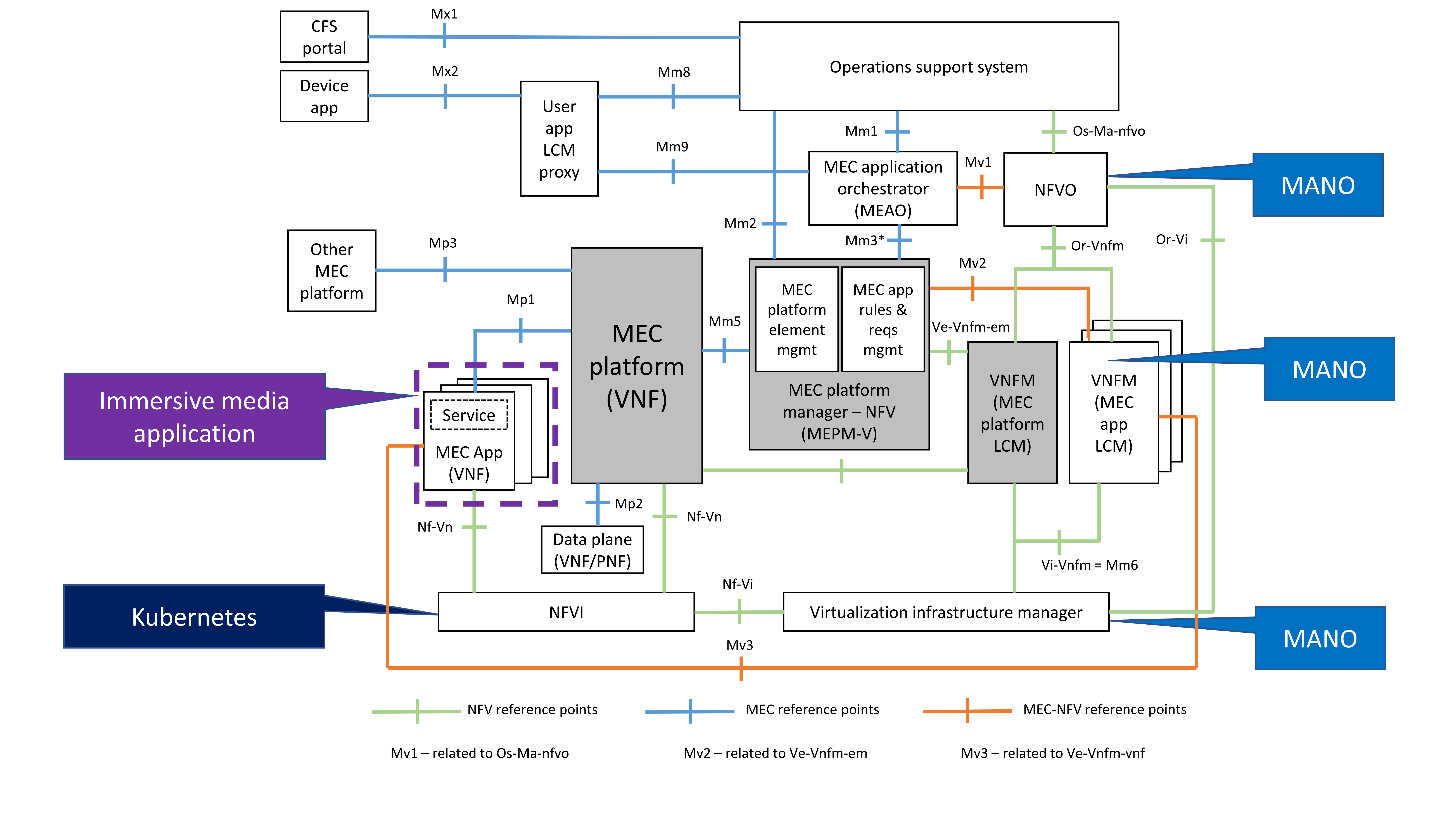}
    \caption{
    Serverless Tele-Immersive Media in 5G MEC
    }\label{MEC-in-NFV-ETSI:fig}
\end{figure*}

Finally, in the context of 5G MEC, a FaaS framework is provided as part of the MEC platform. Figure~\ref{MEC-in-NFV-ETSI:fig} shows the ETSI reference architecture for 5G MEC. In this architecture variant, termed \emph{MEC in NFV}, the application components (serverless functions) are required to be packaged as Virtual Network Functions (VNFs) to be managed by the ETSI Management and Orchestration Stack (MANO) via either a Virtual Network Function Manager (VNFM) or a Network Function Virtualization Orchestrator (NFVO). Finally, the actual container allocation should be performed by Virtual Infrastructure Manager (VIM) MANO component. Therefore, a challenge arises in how to harmonize ETSI MANO standards with FaaS. We partially addressed this problem in our previous work~\cite{AlvarezTBC-2019}, where we described an ETSI compatible FaaS VIM. In this paper, we deal with additional problems related to harmonizing orchestration of serverless functions with ETSI MANO to implement the tele-immersive media application. 

We will now briefly discuss the challenges mentioned above, and outline how we deal with them in the proposed solution\footnote{A reference implementation of our extended FaaS framework and its integration with MANO, can be found in https://github.com/5g-media/faas-vim-plugin.}.

\subsection{Orchestrating Serverless Applications in 5G MEC}
\label{orchestration:subsec}

One of the more important challenges to integration between MANO and serverless technology that we faced in our serverless tele-immersive media implementation was the inability to model a FaaS based service using ETSI VNF Descriptors and VNF Packages. Particularly, a FaaS-based network service includes components that should not be started upon service instantiation, but created and deleted based on custom events. Some of these events are possibly happening inside the application itself. This pattern cannot be reduced to what ETSI MANO already handles well --- auto-scaling. Rather it requires additional flexible orchestration mechanisms, which are application specific. We have developed such serverless orchestration, which generalizes to any custom orchestration scenarios and across multiple use cases. 

\begin{figure*}[!htbp]
    \centering
    \includegraphics[width=\textwidth]{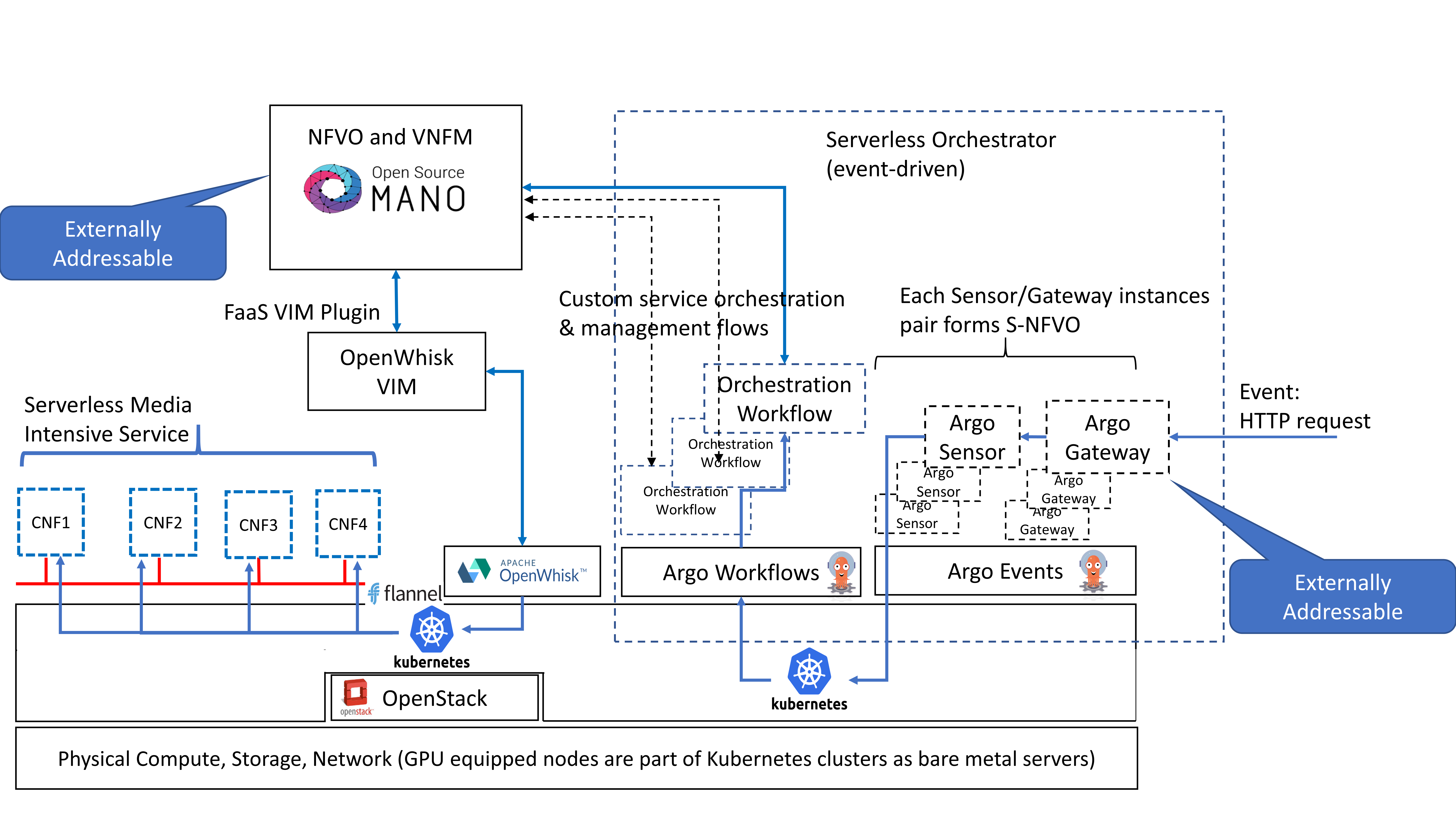}
    \caption{
    Serverless Tele-Immersive Media in 5G MEC
    }\label{media-intensive-fass-in-MEC:fig}
\end{figure*}

In Figure~\ref{media-intensive-fass-in-MEC:fig} we show how we combine serverless orchestration with MANO for the sake of managing serverless tele-immersive media in 5G MEC. We used Kubernetes as our NFVI because of it being a de facto container orchestration standard. Also it provides out-of-the-box capabilities for networking and GPU consumption by the OpenWhisk serverless functions, as we discuss in Subsection~\ref{network:subsec} and Subsection~\ref{gpu:subsec}, respectively.

To harmonize serverless functions with the standard ETSI network service modeling and life cycle management cycle, we add key/value pairs to the optional information field of a Virtual Network Function Descriptor (VNFD), indicating whether a  VNF/CNF is serverless and whether is should be started upon instantiation of the network service or upon some custom event. 

Each serverless VNF/CNF in our system is an OpenWhisk action\footnote{In Apache OpenWhisk parlance, functions are termed \emph{actions}. We will use the terms interchangeably, wherever this does not cause an ambiguity} that is pre-registered with the OpenWhisk FaaS system, which is provided as part of the 5G MEC Platform. This is part of the onboarding into a VIM mechanism prescribed by MANO. For more details see our previous work~\cite{IEEE-Comm-SDK}. The image of a serverless VNF/CNF is simply a fully qualified action name that points to an appropriate metadata associated with the action: key/value pairs describing the action executable, resource limits, such as memory, CPU, maximum execution time, and a variety of annotations that help OpenWhisk to invoke and manage this function. 

This metadata is interpreted by our OpenWhisk VIM that does not invoke an action (does not start the serverless VNF/CNF) upon the initial instatiation driven by ETSI MANO (shown on the left of the figure). A standard MANO instantiation flow consists of un-packaging the network service package, instantiating VNFs/CNFs in the sequential order of appearance of their VNFDs in the package. MANO's NFVO of VNFM polls VIM periodically to obtain metadata on the instantiated VNFs/CNFs when they get started and configured with Day 0 configuration (e.g. IP:port address). This data is  being stored inside MANO's in-memory inventory of the running VNFs/CNFs, called Virtual Network Function Record (VNFR). Our OpenWhisk VIM initially makes up default metadata, such as ``0.0.0.0:0" for serverless VNFs/CNFs that should not be started upon instantiation, but rather upon events.

To handle event-driven instantiation and configuration of serverless functions, we developed a novel orchestration subsystem, which is shown on the right side of Figure~\ref{media-intensive-fass-in-MEC:fig}. We use CNCF Argo Workflows~\cite{argo-wf} and Argo Events~\cite{argo-events} as the basic mechanism for the proposed serverless orchestration. The former is a Kubernetes-native workflow management engine, while the latter is a Kubernetes-native event dependency resolution system that can trigger Argo Workflows in response to external events. We include a special \textit{bootstrap} function  with every network service that uses serverless functions that should be started on demand. In particular, in our implementation of the tele-immersive gaming, transcoders and replay functions are started on demand in response to in-application events rather than upon the initial instantiation. 

The bootstrap function contains \textit{yaml} definitions for two Kubernetes \textit{Custom Resources (CRs)}: Gateway and Sensor, which are specific to this network service. The CRs comprise the standard Kubernetes mechanism to extend the Kubernetes resources ecosystem, so that external resources can be managed like native ones, such as pods, deployments, jobs, etc. An interested reader is referred to the Argo documentation for details of how to use Argo. For the sake of the exposition in this work, it suffices to mention that CRs are essentially yaml files adhering to the Argo dialect. Each such file, a CR, is an instance of a schema called Custom Resource Definition (CRD). Gateway and Sensor lifecycles are managed by Gateway and Sensor controllers, respectively, which watch for the new CR instances of Gateway and Sensor CRDs. When such instances appear as a result of applying the CR document to the Kubernetes API server, the Gateway controller sets up a new Gateway instance and connects it to an external event source and a Sensor target, as specified in the CR specification. Likewise, when a new Sensor CR is applied, a Sensor controller that watches the Sensor CRD creates a new Sensor instance and makes itself available to receiving events from the appropriate gateways.

In our implementation, the Argo Gateway, Sensor, and Workflow controllers are part of the pre-deployed services provided by the MEC (see Figure~\ref{MEC-in-NFV-ETSI:fig}). A bootstrap function is always started upon instantiation and immediately after starting, and applies yaml CR definitions of Gateway and Sensor for this service instance, thus creating a session level event-driven orchestration control plane. This control plane exists for the duration of the service and once the service is deleted (or naturally comes to a termination, e.g. if the game time is up), this event driven control plane is purged from the system. 

Our implementation uses an out-of-the-box Webhook Gateway that can receive external HTTP requests that it passes to the Sensor. The sensor is more intricate. Based on the payload of the HTTP request (i.e. an event that it receives from the Gateway), it conditionally executes lifecycle management actions, such as starting a serverless VNF/CNF, stopping a VNF/CNF, Day2 configuration related actions, etc. A service developer has to program the Sensor to enable this event-driven orchestration at runtime.

The lifecycle management action is an Argo Workflow instance (yet another CRD) that natively executes in Kubernetes. In essence, when a Sensor conditionally triggers Argo workflows dependent on the operation passed to the Sensor by the Gateway. The operation specification is part of the original HTTP request. The triggering is performed by applying a corresponding Argo Workflow CR instance to the Kubernetes API Server. The Argo Workflows controller that watches for the new CR instances pick it up and sees for the workflow execution.  

We use this this novel orchestration mechanism as follows. When our network-centric optimization decides to reallocate specific transcoding profiles, the control plane of the application that performs the optimization of a specific session sends an HTTP request to the Gateway of that session (previously set up by the bootstrap function upon instantiation of the service) requesting termination of some transcoder profiles and allocation of some other profiles (i.e. terminating some running OpenWhisk actions and invoking some other OpenWhisk actions in the Kubernetes NFVI through the OpenWhisk API). Likewise, when an event of interest happens in the session, an HTTP request to start a replay function is sent to the Gateway of the session, triggering a management workflow in the Sensor of the session that invokes the replay action, configures it and connects it to the rest of the running service. 

\subsection{Networking for Serverless Applications in the 5G MEC}
\label{network:subsec}

FaaS frameworks do not support direct network communication among functions out of the box. In our prototypical implementation, we use Kubernetes as the backend for OpenWhisk actions (containers) execution. Kubernetes provides a number of networking solutions out of the box through its Container Network Interface (CNI) standard. These solutions differ in the level of maturity and sophistication. A survey of the Kubernetes networking landscape is out of scope for this paper, so we provide only a high level description pertaining to our implementation. In our proposed solution, we use Flannel~\cite{flannel}, a simple pod level overlay network that can be used to enable containers running in these pods to communicate directly.
The challenge in using Flannel for our work was in devising the orchestration workflows in the Sensor to set up the network just in time upon the service instantiation and then connecting the newly invoked OpenWhisk actions (which eventually run as pods) to get connected to this network. 

A typical hard problem associated with this is port mapping. For each pod in Kubernetes, an IP address of the pod is the address of the Kubernetes Master (also known as an address of the cluster). However, ports should be allocated dynamically and without conflicts. For internally addressable (i.e. within the same Kubernetes cluster), the port mapping is automatically solved by using a NodePort resource that exposes a pod as a service. However, in our case, if a service component should be accessed externally, a more elaborated Ingress resource should be defined. We omit the technical details of setting up and configuring the Ingress resource and Ingress Controller. It is important to stress that in our system this is being done on demand using our serverless orchestration mechanism described in the previous subsection.

\subsection{GPU Allocation for Serverless Applications in 5G MEC}
\label{gpu:subsec}

Some transcoding profiles require GPUs for efficiency. In fact, a large part of this work is devoted to optimizing usage of GPUs for serverless tele-immersive media applications in 5G MEC, where these resources might be scarce and relatively expensive. However, before we can optimize usage of GPUs by serverless frameworks, we need a basic support for consuming them. Apache OpenWhisk proved to be an easily extensible framework in this respect.
OpenWhisk contains an extensible dictionary of action \textit{kinds} that define their runtimes. We created a new runtime that uses NVIDIA's CUDA framework. For example, a generic CUDA action can be defined as follows:

\begin{lstlisting}[caption={CUDA action},label=cuda-action]

"cuda":[
 {
        "kind": "cuda:8@gpu",
        "default": true,
        "image": {
             "prefix": "docker5gmedia",
             "name": "cuda8action",
             "tag": "latest"
        },
        "deprecated": false,
        "attached": {
        "attachmentName": "codefile",
             "attachmentType": "text/plain"
        }
 }
\end{lstlisting}

Adding an entry to the action kinds dictionary is not sufficient to make Apache OpenWhisk to interpret this new action kind. There is a component in Apache OpenWhisk called Kubernetes Client, which --- when OpenWhisk is being configured to use Kubernetes as a container management environment for the actions --- creates a Kubernetes pod yaml definition out of the action metadata. This yaml definition is then applied by the Kubernetes Client to the Kubernetes API Server and the action starts executing as a Kubernetes pod. A Kubernetes yaml definition for the action shown in Listing~\ref{cuda-action} would look as follows:

\begin{lstlisting}[caption={CUDA action Kubernes yaml definition},label=cuda-action-k8s]
apiVersion: v1
kind: Pod
metadata:
  name: cuda8action
spec:
  containers:
    - name: cuda8action
      image: "docker5gmedia/cuda8action:latest"
      resources:
        limits:
          nvidia.com/gpu: 1
\end{lstlisting}

We modified Apache OpenWhisk Kubernetes client to recognize GPU action kinds that we defined for transcoding profiles that use GPUs. When such an action is on-boarded on OpenWhisk that is configured to work with Kubernetes that has GPU equipped worker nodes in its cluster, the action will be placed by the Kubernetes scheduler to a node that has NVIDIA GPU (this functionality is being supported in Kubernetes as an experimental feature since Kubernetes 1.8).
A full implementation is available in~\cite{GPU-OW-blog}.

\section{Serverless Adaptive Streaming Service}
\label{sec:serverless}
In this section, we describe the design of our adaptive serverless streaming service.
In our implementation we used FaaS, as described in Section~\ref{extended-faas}. Although FaaS actions have limited life time, we found them adequate for implementing our short-session-based service.

\textit{Tele-immersive} media streaming services are usually sporadic in nature, with long periods of idleness interspersed with short sessions of activity (e.g. gaming or conferencing). As a result, under a traditional VM-based design, apart from the increased service complexity, a constant sizing problem would manifest when seeking to optimize the service's costs. FaaS offers a more cost-efficient alternative as it automatically scales to the number of active sessions.

As media streaming consumers can have very different bandwidth or processing capabilities and network conditions can fluctuate, a crucial part of an effective media streaming service is adaptation. The original content is transcoded into a number of media profiles, each targeting a different bandwidth and media quality, allowing each consumer to receive the profile most suited to their needs. Lack of an appropriate profile can lead to frequent buffering events, for on-demand consumption, or make meaningful reception completely impossible for live streaming. Hence, adaptation is especially important in live streaming media services.

An apparent advantage of a serverless adaptive media streaming service is more efficient utilization of its available resources, such as different transcoding profiles. Indeed, for smaller consumer size sessions, not all profiles might be relevant, which allows for cost-optimized use of resources. Thus, apart from inter-session scaling, serverless streaming offers the capability of finer-grained intra-session scaling and adaptation.

This is more pronounced for emerging media services, whose profiles and codecs have not yet converged to a standard, in contrast to traditional (i.e. flat/2D) audiovisual media. Thus, emerging media has to deal with a wider repertoire of profiles. Specifically for 3D immersive media~\cite{karakottas2018augmentedvr},~\cite{christaki2019space}, the profile selection problem is more complex~\cite{Doumanoglou_Drakoulis_Zioulis_Zarpalas_Daras_2019} due to the simultaneous availability of various profiles (joint 2D and 3D) and their suitability to highly heterogeneous consumer types (i.e. mobile, workstations, VR headsets) that in turn, come with different requirements in terms of the delivered profiles.

This type of immersive media delivers two payloads simultaneously, the 3D mesh media stream and the multi-view textures media streams.
While the latter are encoded with traditional flat/2D media encoders, the former use distinct 3D codecs.
This effectively renders each immersive media stream profile to be a tuple of a video (i.e. 2D), $P_{2D}$, and 3D, $P_{3D}$ profiles, leading to a more complex visual quality formulation \cite{doumanoglou_quality_2018}.
Furthermore, emerging consumption means that VR and AR accompany traditional displays (i.e. desktop/laptop and mobile), creating a far more complex landscape for profile selection that depends on each consumer type's computing and viewing characteristics.
We argue that for sessions with relatively few consumers, which will require only an optimal subset of profiles, a serverless streaming model is more appropriate, because it opens up more opportunities for optimization.

As explained in the previous section, our extended FaaS framework allows GPU consumption. This adds another dimension to our profiles, expanding the two-tuple to a three-tuple $(P_{3D}^a, P_{2D}^b, R^c)$, containing the 3D and video profiles, in addition to the computing resource type $R^c$ (i.e. CPU: $c=0$; or GPU: $c=1$).
Thus, profiles with similar bit-rates, may reduce processing latency at the expense of using higher cost resources.
For conciseness, we denote a transcoder's joint 3D media profile as $P^n$, with $n$ encoding a unique combination of $a, b$ and $c$.

Serverless design follows a single responsibility principle: each function is responsible only for a single task, instantiated as the need arises and destroyed when the task is completed. In the context of media streaming adaptation, this translates to having one transcoding function for every combination of profile and source (i.e. a player).

A general scenario for tele-immersive media streaming includes a population of producers ($K$), which generate live 3D video streams; and a population of spectators ($S$) who need to receive the streams of all producers and reconstruct them in the virtual environment. Our service then comprises a broker function (\textit{vBroker}) and $|N| \cdot |K|$ transcoder functions (\textit{vTranscoder}), where $N$ is the set of transcoding profiles, as each vTranscoder is responsible for transcoding the stream of one specific player to one specific profile.

Producers send their production streams to the vBroker, while vTranscoders receive these streams from the vBroker, transcode them according to predefined profiles, and upload them back to the vBroker. Consumers then are served either the production stream or a transcoded one from each producer, based on an adaptation logic. In the context of Kubernetes, this means allocating a set of transcoding actions for each production stream.

This serverless adaptive streaming design lends itself to optimization. Transcoder functions can be deployed on demand while monitoring the service's behaviour, as events in response to the monitoring analysis.
Typically this relates to monitoring its cost, and seeking to minimize it, and monitoring the QoE of its consumers, seeking to maximize it.
Taking into account the cloud-native transformation happening thanks to the emergence of 5G and the virtualization and softwarization of the network, it is possible to perform service optimization in an integrated manner with the network itself.
Instead of relying on a local client-based adaptation, service adaptation and optimization can take a more global approach.

Our streaming service is entirely dynamic, with the vBroker action deployed at the start of each session, for that specific session. 
This allows for edge proximity placements and a flexible vBroker interconnection scheme that unifies edge and core resources, allowing our session based services to span multiple infrastructures.
Transcoders are deployed on-the-fly according to the network-centric session optimization logic.
The service has a choice either starting with zero transcoders and subsequently adding them on demand as guided by the optimization, or starting with a default transcoder profiles configuration, and then adapt it to the actual consumers. This is similar to client-based adaptation that start either on the lowest/highest profile, and then adapt to that which results in higher QoE.


Application specific events (e.g. replaying highlights) trigger processing functions that are deployed on the serverless infrastructure and are responsible to synchronize media and game-state streams to produce replay clips that they can later be served to spectators on demand.

\begin{figure*}[!htbp]
    \centering
    \includegraphics[width=\textwidth]{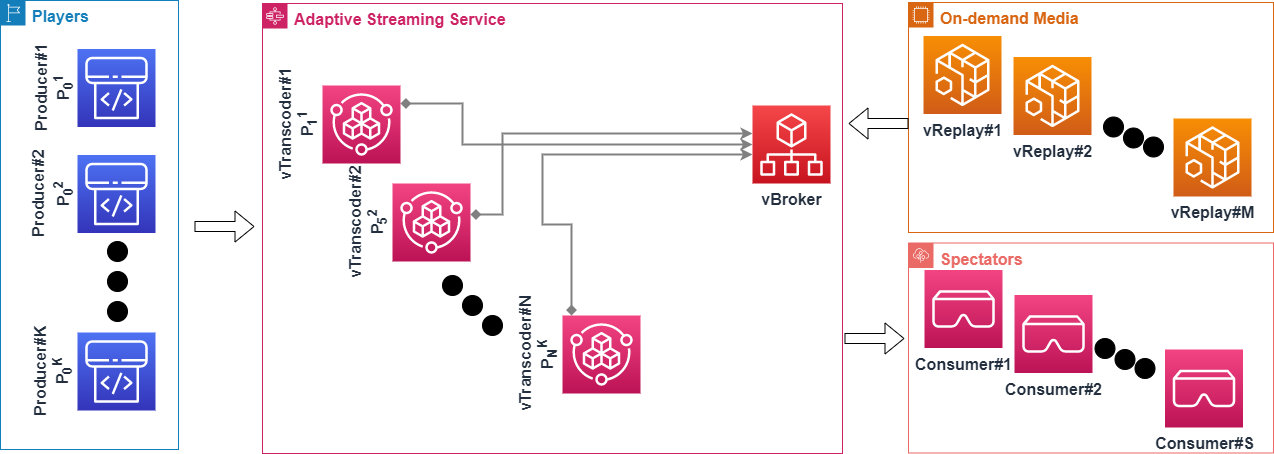}
    \caption{
    An abstracted architecture of our service: vTranscoders are instantiated or destroyed as needed, each one responsible for transcoding the media stream of one producer to one profile. vReplay functions are similarly triggered by certain events. All streams flow through the vBroker, from which consumers receive the allocated media streams.
    }\label{fig:service_compoenents}
\end{figure*}

In Fig.~\ref{fig:service_compoenents}, the service components are depicted. 
On the left, producers in the 3D immersive media production platform produce high quality profile 3D media streams, denoted as $P_0^{k}$. 
The adaptive streaming components are comprised of a set of vTranscoders each one being responsible for transcoding an input 3D media stream from a single player to a single profile. 
Those transcoded streams become available to the consumers via the vBroker instance. 
Additionally, 
vReplay instances are instantiated on the FaaS infrastructure in response to specific events, as described in Section~\ref{extended-faas}.
Upon the completion of replay clip processing, the processed media clips become available to the application consumers on demand.

In more detail, our network-centric real-time adaptive streaming service drives an  AugmentedVR \cite{karakottas2018augmentedvr} gaming application.
The application manages gaming sessions supporting $K$ players and $S$ spectators, where $|S| \gg |K|$.
Each player is captured with a volumetric capturing station \cite{sterzentsenko2018low} and 3D-reconstructed in real time \cite{alexiadis_integrated_2017}, producing a live 3D media stream.
The players' live media traffic, along with the application game state metadata are transmitted and synchronized among the playing users (more details regarding the application's architecture can be found at \cite{christaki2019space}).
In this way, players are emplaced within the same shared virtual environment, and interact within it under a capture-the-flag context.
Through the aforementioned adaptive streaming service back-end, the application allows for remote party spectating of each gaming session.

\begin{figure*}[!htbp]
    \centering
  \includegraphics[width=0.85\textwidth]{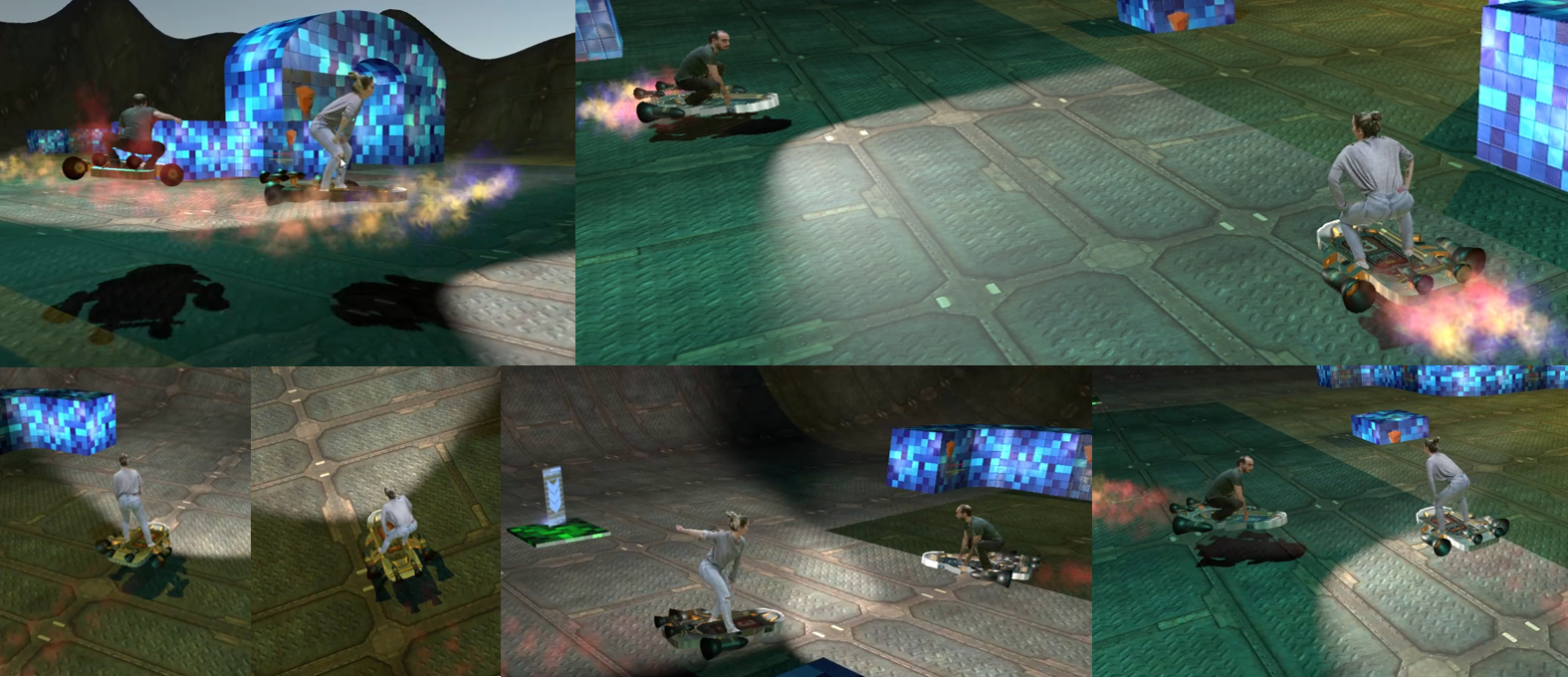}
    \caption{
    Screenshots of the AugmentedVR \cite{karakottas2018augmentedvr} immersive media game where the playing users real-time 3D media streams are embedded into the same shared virtual environment.
    The screenshots' viewpoints are those of spectating users that can freely navigate the scene in order to spectate the action around the virtual arena.
    }
    \label{fig:app_screenshots}
\end{figure*}

The spectators $S$ receive the synchronized game state and all $|K|$ players' media streams, faithfully reproducing the current session, with example screenshots presented in Fig.~\ref{fig:app_screenshots}.
While the players' communication is based on stringent real-time requirements, the spectators' media consumption relies on broadcast traffic, and thus requires consistent streaming with relaxed latency constraints.
This is driven by a centralized control plane of the application, which oversees the production and delivery of appropriate profiles to each spectator for smooth playback. The control plane is extensible and new optimization algorithms can be plugged in as needed. In Section~\ref{subsec:optimization} we present our proposed network-centric optimization to drive the control plane and in Section~\ref{sec:experiments} we compare this smart optimization with a more naive baseline algorithm to quantify the benefits of the network-centric optimization. The control plane of the application interacts with the Serverless Orchestration mechanism described in Section~\ref{extended-faas} to actuate the transcoder profile allocation plans calculated via optimization. 

These profiles are selected from a set of profiles $N$, with each spectator receiving one profile $P^\mathit{k}_\mathit{n} \in N$ (with $n \in |N|$) for each player $k \in K$. Each profile is served by a single transcoding action, spawned and managed by the service, that re-encodes the originally produced profile $P^\mathit{k}_\mathit{0}$ from a specific playing user, to a lower bit-rate profile $P^\mathit{k}_\mathit{n}$, which is made available on the broker.
At the same time, the application orchestrates the production of on-demand media in the form of highlight replay clips.
These are event-driven processing actions that produce finite media streams of previously captured live traffic.
Once produced, these too are available on the service's broker for on-demand consumption by the spectators.
Finally, the orchestration and management of the transcoding actions are handled by our service's optimization logic that has a dual role.
On one hand, to optimize the application's costs, while preserving the resulting QoE by making scaling decisions for its elastic components (i.e. the transcoding actions);
and, on the other hand, to apply network-centric adaptation by collectively deciding each spectator's consumed profile.

One important design concern is dealing with the fixed maximal life time of FaaS executions. In cases, when the session time is about to exceed the lifetime of the functions involved, a shadow FaaS invocations can be started and configured. As explained in Section~\ref{extended-faas} we use NodePort to expose serverless functions as Kubernetes services. This means that we can transparently switch one FaaS invocation by another without disturbing the service. Therefore, while any concrete serverless function cannot execute beyond its maximal life time, collectively an intensive media session can be extended as needed at fine granularity.


\section{Network-centric Cost Optimization}
\label{sec:formulation}
When considering the optimization of a serverless live streaming delivery network, there are two conflicting objectives: to maximize the QoE of every individual spectator and to minimize the cost to the service provider. Maximizing the QoE entails making the streams available in multiple versions differentiated in visual quality and bitrate, so that each spectator can consume a version most suited to her device type, processing power and connection capabilities. The production of multiple transcoding profiles, however, involves running more transcoder FaaS functions, thus increasing cost.

In order to balance a tradeoff between QoE and cost, both must be expressed in common units. Providing a certain QoE level can be naturally connected to generating revenue for the stream producer, either directly or indirectly. Our proposed optimization maximizes profit for the stream provider (i.e. the revenue minus cost objective). This section describes the components involved in modelling revenue and cost.

  

\subsection{Spectator behavior}
\label{subsec:spectator-behavior}

During the course of a session of live-streamed media, individual spectators may be consuming the stream from its start, or join at any later point in time. Streams of different characteristics (e.g. popularity) may attract new spectators at different rates and numbers. Similarly, spectators may stay online until the stream ends, or quit before that, for reasons which may or may not be relevant to the stream characteristics.

\subsubsection{Spectator arrival} \label{Joining}

Traditionally, an arrival process of people to stores, facilities, telephone calls have been modeled using Poisson distribution~\cite{Kleinrock1976},~\cite{jain1990art}. The Poisson distribution calculates the probability of \(k\) events (e.g. arrivals) occurring in a specified interval, given the average number \(\lambda\) of events per interval  \cite{yates2014probability}:

\begin{equation}
  P(k \text{ events occurring in interval}) = \frac{\lambda^k e^{-\lambda}}{k!}
  \label{eq:Poisson}
\end{equation}

The process of spectators arrival to an immersive gaming session involves humans making discrete decisions on joining a gaming session. Modeling spectators' arrival to immersive gaming requires a thorough study and careful characterization. However, we are unaware of the literature presenting accurate stochastic modeling for spectators' arrival. Probably, this can be explained by this domain being relatively new and rapidly evolving.

It is reasonable to assume that in certain settings the overall arrival process in immersive gaming might differ considerably from the Poisson distribution. For example, a large number of spectators can arrive simultaneously at the beginning of a session if a session is perceived as extremely interesting because of the prominence and high ratings of the players. More spectators might arrive later and their arrival process can follow the Poisson distribution with the overall distribution being non-Poisson. Note that Poisson inter-arrival process requires that the inter-arrival times are distributed exponentially. Obviously, a simultaneous bulk joining of spectators breaks this assumption. Likewise, a session inter-arrival process can start as a Poisson one and then a bulk of spectators can join all at once, e.g. because of a viral notification in a social network that this session is a must to attend. Obviously, the Poisson assumptions will be broken in this case as well.  

These are extreme scenarios which, albeit possible, are not necessarily expected in each and every gaming session. Furthermore, these scenarios are not amenable to the fine grain optimization that we propose in this paper, because statistically in large spectator populations we would expect to have enough members with every possible terminal capability to justify allocation of all possible combinations of transcoding profiles to maximize the total QoE for spectators. This would make the proposed serverless architecture to perform no better than other approaches.

To explore and characterise the sweet spot of our proposed optimization, we focus on relatively small sessions in terms of spectators number and make an assumption that these spectators will behave similarly to the viewers of the video streaming services. It should be noted that spectators consume video and because the sessions are relatively short and dynamic their behavior might be similar to that of the video streaming services customers browsing video content and watching previews. It is widely accepted to model the inter-arrival process of such customers using Poisson distribution~\cite{JinGISMO-2001,Bentaleb_DQ_DASH2020} even though some works exist that indicate that the inter-arrival process to the media streaming services can be better modeled by the lognormal distribution~\cite{Ali_Eldin2015}.

In this work, we decided to follow the mainstream approach and model inter-arrival process of spectators to the gaming sessions as a Poisson distribution. We define the distribution's interval as the ten-second time step  and set the average number \(\lambda\) of arrivals per interval  to a range of values from 0.25 to 1, with a default of 0.5 arrivals per ten-second time step. 
\ignore{
\unsure{David: Kostis, please add the numerical specifics of your modeling here}
}

\subsubsection{Spectator quitting} \label{Quitting}

Once spectators join, they may remain online until the end of the stream or quit before then. 
Chen et al. \cite{chen2015qoe} model spectator quitting probability as a function of their QoE: a spectator with very bad QoE is certain to quit, while a spectator with very good QoE is likely to remain but still has some 20\% probability of quitting before the session ends, for non-QoE related reasons. 
Between these two extremes, the decrease of quitting probability with QoE is assumed to be linear.

In a scenario with a diverse mix of spectators, QoE may vary significantly depending on device type, processing power and connection bandwidth. 
Spectators with powerful PCs and a good connection will have a better QoE than spectators with mobile devices, which would lead to mobile spectators quitting much more frequently. 
In this work we consider that each spectator is aware of their own hardware and connection capabilities, and will be happy with the best QoE possible for that configuration. 
Hence, in estimating quitting probability, we consider the \emph{difference} between the maximum QoE possible for each spectator and their actual QoE. 

Other factors that might impact quitting for non-QoE related reasons include the interest level of a given session: spectators may abandon a boring or slow session more easily than a very active or thrilling session. 
This will also impact QoE-based quitting probability modeling, as spectators may be reluctant to leave an interesting stream despite QoE being mediocre. QoE-related quitting can be further altered by how demanding a spectator population is.

Hence, based on the findings of \cite{chen2015qoe} and these considerations, we build a linear quitting model for each 10-second time step. The probability that during a time step \(t\) a spectator experiencing  \(QoE_t\), will quit, is:

\begin{equation}
  q_t = q(QoE_t) = b + (QoE_{max} - QoE_t) \cdot d = b + dQoE \cdot d
  \label{eq:quitting-prob}
\end{equation}

\noindent where:
\begin{itemize}
    \item \(b\) is the base quitting probability per time step for non-QoE related reasons, with a default value of 0.37\%, corresponding to a cumulative probability of 20\% to quit at some point in the course of a 10-minute session. 
    \item \(dQoE^s\) is the difference from the maximum possible QoE for that spectator.
    \item \(d\) is a factor denoting how much QoE impacts quitting, which is dependent on QoE value range produced by the QoE model and the session parameters (i.e. how interesting or important a session is, and consequently how likely spectators are to leave because of QoE dissatisfaction). The QoE model we adopt (see \ref{QoE_model}) produces values usually within the range of 2.8--3.8. Hence, \(d\) ranges from 10\% (an interesting session that spectators won’t quit easily) to 50\% (very demanding spectators), with a default value of 20\%.
\end{itemize}

The probability that a spectator remains online in a given time step is \(p_t = 1 - q_t\). The probability of a spectator to remain online from \(t_0\) to \(t_1\) would be the product of remaining at each individual time step in between, which, naturally, is decreasing over time:

\begin{equation}
  p_{t_0 \rightarrow t_1} = \prod_{t=t_0}^{t_1} p_{t} = \prod_{t=t_0}^{t_1} (1 - q_{t})
  \label{eq:cum_quitting-prob}
\end{equation}

Eq. \ref{eq:cum_quitting-prob} assumes that quitting events during different time steps are independent and identically distributed. While this may not always be the case, a more sophisticated model of spectator behavior is currently outside of the scope of this work, because more field data should be collected on immersive media spectators online behavior as these services become mainstream. Presently, this is still a new area and we believe that using simpler modelling is justified forr initial exploration of cost/QoE trade-offs.  

To calculate the probability of a spectator remaining active from the beginning of the session to its end, \(t_0\) and \(t_1\) can be set to 0 and \(|T|\), respectively. For a 10-minute session comprised of 10-second time steps, \(|T|\) would equal 60.

Hence, for example, a demanding spectator in a boring session, with a \(dQoE\) of 0.5, might have a 5.37\% probability to quit every 10 seconds, meaning she may soon leave unless her QoE improves. Note that in the relatively narrow QoE range produced by the QoE model (see Section \ref{QoE_model}), a \(dQoE\) of 0.5 represents a significant decrease from the optimal QoE for this spectator.
Conversely, for an undemanding spectator in an interesting session with a \(dQoE\) of only 0.1, quitting probability would be 0.57\% per ten-second time step, and he is 71\% likely to remain until the end of a 10 min session.

\subsection{Revenue}
\label{subsec:revenue}
\ignore{David: repetitive, omitted.
To recover production and delivery costs and also motivate offering of the streaming process, we assume that spectators generate some revenue for the producer. 
}
Depending on the use-case and the marketing approach, revenue for the media stream service provider can range from direct (e.g. a subscription-based or pay per-use service) to indirect (e.g. a service supported by ads).

In general, the provider is interested to keep spectators engaged for longer time periods, because it might generate more revenue. In an ad-supported service, spending more time watching the stream results in greater exposure to the advertisements. In a subscription service, spectators who don't spend so much time watching the stream may reconsider renewing their subscription. Spectator QoE may also impact the revenue they are generating, or not, depending on the specific use-case. In a pay per-use service, the revenue generated is directly proportional to the time spent in the service.

In this work, we consider an ad-supported use-case as a baseline scenario, and correspondingly assume that each active spectator generates indirect revenue per time unit, so long as they remain active. Revenue generated per time step can be constant, or a function of the spectator's QoE, considering that spectators happier with their QoE may be more receptive to ads. As revenue modelling varies by use-case and is outside the scope of this paper, we consider the generic case that revenue is a function of QoE. This can be modelled by any monotonically non-decreasing function, e.g. constant, linear or logistic:

\begin{align}
       &r_t=r(QoE_t) = \vphantom{\frac11}a_1    &\text{(constant)} \label{eq:revenue-constant}\\
       \text{or}\hspace{10mm} &r_t=r(QoE_t) = \vphantom{\frac11}a_2 \cdot QoE_t  &\text{(linear)} \label{eq:revenue-linear}\\
       \text{or}\hspace{10mm} &r_t=r(QoE_t) = a_3  \cdot \dfrac{1}{1+e^{-QoE_t}} 
       &\text{(logistic)}  \label{eq:revenue-logistic}
\end{align}

Over the course of a streaming session, the revenue generated by a spectator during each time step accumulates to produce the total revenue over time:

\begin{equation}
    r_{t_{join} \rightarrow t_{quit} } = \sum_{t=t_{join}}^{t_{quit}} r_t
\end{equation}

However, the time that spectators remain active, and therefore generate revenue, is directly affected by the QoE they are experiencing, as mentioned in subsection \ref{Quitting} and Eq. \ref{eq:quitting-prob}. For a given future time step \(t\), the \emph{average expected} revenue generated by a spectator with \(QoE_t\) and \(q_t\) probability of quitting will be dependent on the probability they remain active until \(t\). Taking into consideration Eq. \ref{eq:cum_quitting-prob}:

\begin{equation}
        E(r_t) = r_t \cdot p_{t_0 \rightarrow t_1}  \\
        =  r_t \cdot \prod_{t=t_0}^{t_1} (1 - q_{t})
    \label{eq:expected_revenue}
\end{equation}

Therefore, taking into consideration Eq \ref{eq:expected_revenue} and the dependency of \(r\) and \(q\) on QoE for every time step, the total expected revenue from a spectator, from the current time \(t_0\) until time \(t_1\) is:

\begin{equation}
    \begin{split}
        E(r_{t_0 \rightarrow t_1}) &= \sum_{k=t_0}^{t_1} E(r_k )
        = \sum_{k=t_0}^{t_1} \left[r_k \cdot \prod_{t=t_0}^{k} (1 - q_{t}) \right] \\
        & = \sum_{k=t_0}^{t_1} \left[r(QoE_k) \cdot \prod_{t=t_0}^{k} \Big(1 - q(QoE_t)\Big) \right] 
    \end{split}
    \label{eq:cumulative_revenue}
\end{equation}

Eq. \ref{eq:cumulative_revenue} highlights how QoE can impact revenue both directly, by altering the revenue an active spectator generates per time unit, and indirectly, affecting their probability of quitting early.

\subsection{QoE model} \label{QoE_model}
In order to keep spectators from quitting the stream early, thus maximizing generated revenue, an optimization algorithm would need to know what each spectator's QoE is at present, and how it may change depending on the \ignore{CNO's} network-centric optimization decisions.
Although a number of video streaming QoE models exist (e.g. \cite{paudyal_impact_2016,robitza2017modular,xin2019user}), there is none, to our best knowledge, that regards textured 3D meshes viewed in a free viewpoint environment.
However, for testing purposes, a suitable 2D video QoE model may be adopted.

In this paper we derive our QoE model from Zadtootaghaj et al.~\cite{zadtootaghaj2018modeling}. 
In that work the authors consider cloud gaming, which is a close match to our own use-case. 
Using subjective mean opinion score (MOS) measurements, they derive QoE as a second degree function of image PSNR and frame rate (FR), fitted to the MOS:

\begin{equation}
    \begin{split}
        QoE = &-8.97 + 0.056 \cdot \text{FR} + 0.41 \cdot \text{PSNR} - 0.0038 \cdot \text{PSNR}^2 \\
        & - 0.001 \cdot \text{FR} ^2 + 0.00079 \cdot \text{FR}  \cdot \text{PSNR}
    \end{split}
    \label{eq:qoe}
\end{equation}

Knowing the average PSNR and frame size for each transcoding profile and each spectator's bandwidth, we use this model to calculate each spectator's QoE at present and estimate their QoE in the future for different profiles.

In a tele-immersive game, a spectator will be receiving each player's 3D representation in a transcoding profile. For each profile, the average PSNR is known, calculated from the PSNR of the textures used to color the 3D mesh, considering that part of the screen occupied by the 3D reconstruction. Although the latter of course varies by a spectator viewpoint, in the vast majority of cases the 3D reconstruction will occupy an area of 1-5\% of the total screen area. Given that the area not occupied by the 3D reconstruction is computer-generated and suffers no loss of quality with different transcoding profiles, we offset average texture PSNR to obtain an estimate of average spectator view PSNR.

Depending on a spectator's maximum bandwidth, they may be unable to receive the incoming stream at its full framerate. Eq. \ref{eq:qoe} considers the actual framerate experienced by a spectator, which will be dependent on that spectator's connection bandwidth and the average frame size of the received profile.

In the general tele-immersive scenario, each spectator will receive transcoded media streams from \(|K|\) players. Each received stream might employ a different transcoding profile and have different PSNR and framerate values, thus resulting in a different QoE, regarding the quality of the 3D reconstruction of a particular player.

The total QoE for each spectator, which aims to reflect their satisfaction with the whole immersive experience, will be a function of the individual QoEs corresponding to each player. A simple approach would be to simply average the QoE of each player's 3D reconstruction. A more thorough modelling, which is beyond the scope of the present work, might take into account the position and orientation of the spectator and the players inside the virtual space and assign greater weights to the 3D representations of players closer to the spectator, and nearer the center of their field of view. However, position and orientation would not likely remain constant in an immersive environment, even for a ten-second time step.

In this work we opt for the simple averaging approach, assuming that spectators can see both players equally in the virtual space. This in no way limits the generality of the methodology and outcomes, as it considers the most generic case. 



\subsection{Costs}
\label{subsec:costs}

The costs of delivering live media to a population of spectators are comprised of two separate categories: the cost of running the necessary software to transcode and buffer the data, and the cost of delivering the data to the consumers.

In the serverless approach we examine in this work, each transcoded media quality is being produced by its own dedicated FaaS transcoder. 
We assume that such transcoders are being deployed in a 5G MEC FaaS (e.g. using our extended FaaS framework). Since MEC is, essentially, a cloud deployed at the edge (also referred to by telcos as a \emph{cloud edge}) the business model is similar to that of the cloud, but the resources are more scarce and therefore are likely to be priced differently. Applications (such as our tele-immersive gaming) rent these resources on a pay as you go basis. In fact, the application deployment can be more sophisticated. For example, some low end transcoders for spectators who also can tolerate slower (and thus cheaper) network connections can be be placed in a public cloud, such as those provided by IBM, Amazon, Google, etc. Some other, more demanding transcoders, will be placed closer to the spectators, i.e. in the MEC and utilize the unprecedented 5G connectivity speed and bandwidth at a higher price point, striving at the overall profit maximization for the service provider.

\ignore{
a commercial cloud service (such as IBM, Amazon, Google, etc), and are being billed by the second, according to the memory they require. 
}
In addition to the regular resources available to FaaS in the current commercial offerings, our video transcoders may require the use of a GPU for real-time processing, which will incur additional costs, as described in the following subsection.

In addition to the transcoders, a broker function, active throughout the session, is also necessary to facilitate the media stream traffic. The core broker function also facilitates communication between the players. Therefore it will always be placed in the MEC. 

In cloud-deployed functions, only outgoing (not internal) traffic is usually charged, with typical prices ranging from \$ 0.05 to \$ 0.10 per GB. However, the Cloud cannot match KPIs In 5G MEC deployments it is too early to reason about the pricing plans for inside-edge traffic and in-bound and out-bound traffic between 5G MEC and the cloud, because commercial offerings are just being formulated. In general, there are two pricing approaches for the 5G traffic: unlimited plans (cheaper if the network is used a lot) and limited ones, which can be quite expensive.

For simplicity, in our experimentation we focus on the model, in which all application functions (transcoders, brokers, buffering, replay) are deployed at the 5G MEC and exclude traffic pricing from the quantitative analysis, focusing on other resources needed for production. However, it should be stressed that our optimization problem formulation is general and includes traffic costs as part of the overall cost modelling.

\ignore{
However, in the small sessions this paper focuses on traffic charges are relatively insignificant. Therefore, even though we include them into modeling for the sake of completeness, we ignore them in our experimentation for the sake of simplicity and better focus.
}

\subsubsection{GPU Pricing Model}\label{pricing:sec}

Over the last few years, GPUs have become essential to a multitude of applications. Cloud vendors have recognized this market potential and have started providing new virtual server families that include GPUs. However, GPUs present some new issues. In particular, GPUs are not easily amenable to sharing among different workloads. This dictates a time-sharing approach and drives up the cost of the cloud based GPU servers. 

Limitations to GPU sharing are especially challenging for serverless computing. If time-sharing is used, then only one serverless function consuming GPU can run at a GPU-equipped virtual server at a time, with the rest of the server resources (CPU, RAM) being wasted. As we go to press, we are not aware of any commercial offering for serverless computing with GPUs. This does not preclude such offerings in the near future as GPU sharing improves (Nvidia\footnote{\href{https://www.nvidia.com/en-us/data-center/virtual-gpu-technology/}{https://www.nvidia.com/en-us/data-center/virtual-gpu-technology/}}, Nuweba\footnote{\href{https://www.nuweba.com/}{https://www.nuweba.com/}}). Furthermore, we believe that a significant progress with building commercial cloud offerings for serverless GPUs will only become possible when shareable GPU architectures will become ubiquitous and this programming model will be consumable at the application level.

In our previous work, we developed a first-of-its-kind prototype for using GPUs with serverless functions. Our prototype uses Apache OpenWhisk and Kubernetes~\cite{GPU-OW-blog}. To enable quantitative reasoning about using serverless computing for tele-immersive gaming in the 5G MEC's FaaS, we need to develop an estimation of a realistic pricing model for GPUs usage in serverless computing. Since the MEC business model is essentially the same as the public cloud business model with an important distinction of resources being more scarce in MEC, which justifies their higher pricing than in a typical public cloud. Essentially, the supposition of MEC is hat it behaves like a cloud in the edge allowing to leverage proximity to users and higher KPIs at possibly higher price points for providers, but overall making more profit by enabling new application capabilities and providing much better QoE that would attract a large customer base. 

We therefore derive our hypothetical pricing plan for MEC using public clouds as a starting point. To that end, we consider a typical CPU-based cloud functions pricing, and CPU-based virtual server pricing vs GPU-based virtual server pricing and develop a speculative model for the GPU based serverless costs. It should be stressed that our intention is neither to propose an actual pricing model for GPU-based serverless computing nor to argue that the profit margins should necessarily be the same as for the CPU-based one. Rather, our intention is to provide an educated guess for what this model might look like and use it to study the pros and cons of our proposed approach quantitatively. 

Our methodology is to assume the same profit margins ratio between the GPU- and CPU-based serverless computing as between GPU- and CPU-based \emph{virtual servers}. The latter is directly observable from the publicly advertised cloud vendors pricing plans. Note that while this assumption can deviate from the actual ratios in practice, a proportionality between the internal cost of production and the profit should exist. Hence, as long as we preserve the directly observable ratios in our estimations, they should serve as a reasonable proxy.

As an example pricing reference point, we consider pricing plans for IBM Cloud. Similar results can be obtained for other cloud vendors. Functions\footnote{\href{https://cloud.ibm.com/functions/learn/pricing}{https://cloud.ibm.com/functions/learn/pricing}} and IBM Cloud Virtual Server Instances\footnote{\href{https://www.ibm.com/cloud/virtual-servers/pricing/}{https://www.ibm.com/cloud/virtual-servers/pricing/}}. ACL1.8x60 and M1.8x64 are the two models of virtual servers with and without GPU, respectively. These two models have the same number of CPUs (8) and approximately the same amount of RAM (60 and 64, respectively). Billing is being done on a monthly basis. At full time utilization (i.e. $720$ hours per month up time), M1.8x64 costs \$$362.88$ at hourly rate \$$0.504$ while ACL1.8x60 costs \$$1402.56$. This means that leasing a GPU-enabled server with other parameters being equal to a CPU-based one is about of $3.8$ times more expensive. Note that what is important in this study is the internal cost.

With the time shared GPU-based serverless computing, the server can run only one GPU based function at a time. Typically, GPU-enabled servers are large. Therefore, running a single GPU-based function is tantamount to fully occupying a large server for the duration of the function lifetime. The number of $15$ minutes long serverless functions per month per server will be $2,880 = 720 \cdot 4$. Therefore, the cost of a single $15$ min execution can be assumed to be \$$0.487 = \frac{1,402.56}{2,880}$. 

To verify this calculation, one can observe that exactly the same number can be obtain by simply dividing the hourly rate of ACL1.8x60 (\$$1.9472$) by $4$ (number of $15$ minute long functions per hour). This would give a base rate of \$$0.00054 = \frac{1.94}{3,600}$ GPU seconds (we assume the same usage of RAM as for the CPU case). 

Note that while running a GPU-based serverless function, the same host can be used to also run CPU-based functions. Otherwise the CPUs and RAM of the GPU based host will be just wasted. As we observed above, the cost ratio between a CPU- and a GPU-based VM is $3.8$. IBM Cloud Functions are being priced at the base rate of \$$0.000017$ per second of execution, per GB of memory allocated (we abbreviate this to per GB seconds). This implies a base rate of \$$0.000064 = 3.8 \cdot 0.000017$ per GB seconds for CPU based functions (when running on a GPU-enabled host). Of course, it is unreasonable that a CPU function will become more expensive in the public cloud just because we introduced GPU-based functions. This means that to keep the CPU functions at the current base rate, GPU-based functions should be made even more expensive, which will increase the ratio between the GPU and CPU serverless computing costs beyond $3.8$ (alternatively, GPU sharing architecture should be developed and deployed to reduce the GPU price when consumed via serverless functions). However, in the 5G MEC, the users can be more receptive to higher price points, because it is expected for the MEC resources to be scarcer and, therefore, more expensive.

A detailed pricing modeling for the time-shared GPU model is outside of the scope of this paper. 
For the sake of modeling costs of serverless GPU functions in this work, we assume that the FaaS is provided on top of the GPU-enabled servers, similar to, say, ACL1.8x60 with the base rates of \$$0.000064$ per GB seconds for CPU-based functions and \$$0.00054$ per GB seconds for GPU-based functions (i.e. an order of magnitude difference in the cost). With this choice, we will be able to avoid inflating the estimated benefits of our proposal while still be able to demonstrate its usefulness.

\subsection{Optimization}
\label{subsec:optimization}


Our goal is to maximize the profit that an immersive game provider accrues from offering the service on the 5G MEC using FaaS. While there are multiple costs involved with provisioning (e.g. storage for replay clip files, databases for managing service subscription, monitoring subsystem, FaaS charges for replay clips, etc.), in this paper, we focus on minimizing the overall payment for serverless transcoders allocated to spectators to maximize their QoE.  

The revenue is assumed to be generated by active spectators, who have a greater probability to remain active if they experience a QoE that is maximally matching their capabilities in terms of their terminal and bandwidth, as discussed in Subsection~\ref{subsec:revenue}.

Higer QoE is produced by the transcoding profiles that consume more resources and, therefore, are more expensive. Since our goal is to maximize profit for the provider, the network-centric optimization should serve spectators a better QoE only if this increase in QoE is expected to produce revenue that exceeds the cost. Conversely, worsening QoE to save costs is justified only when this does not impact revenue too much by triggering too many spectators to quit the stream.

In the course of our network-centring cost-efficiency optimization two sets of decisions must be taken \emph{on-line} based on the metrics reported by active spectators and models for spectator behavior, cost, revenue and QoE developed in previous subsections:

\begin{enumerate}
    \item Which transcoding profiles should be deployed in production at each point in time to minimize costs of production?
    \item Which of the transcoding profiles for each player should be allocated by the service provider to each spectator to maximize their QoE, thus, maximizing revenue?
\end{enumerate}

\ignore{
Achieving cost efficiency depends on an accurate modelling of the costs and revenues. The former depends on the available cloud and 5G MEC commercial offerings for FaaS. The latter depends on the spectators' behavior. The modeling approach of Subsections~\ref{subsec:spectator-behavior}--\ref{subsec:costs} is relatively simple and generic, developed with the use-case of immersive 3D media live streaming in mind. Naturally, each use-case will have its own peculiarities, which will need to be modelled accurately and possibly fine-tuned using real data.
}

We now define our optimization problem more rigorously. Table~\ref{tab:notations} summarizes the notations that we use in problem formulation.

\begin{table*}[!htbp]
  \caption{Notation Summary}
  \label{tab:notations}
	\centering
  \begin{tabular}{l | l}
  \textbf{Notation} & \textbf{Description}\\
  \hline \hline
  \emph{Sets} & \emph{Description} \\
  \hline
  $\boldsymbol{K}$  & playing users (players), $k \in \boldsymbol{K}$ \\
  $\boldsymbol{S}$  & spectators $s \in \boldsymbol{S}$; $S_{t}$ is the set of spectators at time $t$ \\
  $\boldsymbol{N}$ & transcoding profiles, $n \in \boldsymbol{N}$ \\
  
  \hline \hline
  \emph{Sequences} & \\
  \hline
  $\boldsymbol{T} = \{t_{i}\}_{i=0}^{|T|}$ & equidistant time steps $t_{0}, t_{1}, \ldots, |\boldsymbol{T}|$, where $|\boldsymbol{T}|$ is the maximal session lifetime \\
  \hline \hline
  \emph{Parameters} & \\
  \hline \\
  $q^{s}(QoE) = q^s_t$       & an estimated probability that spectator $s$ quits at time $t$ for a given $QoE$ value (See Eq.~\ref{eq:quitting-prob}) \\
  
  $p^s_t = 1-q^s_t$        & an estimated probability that spectator $s$ stays in the session at time $t$ \\
  $\vec{n}= (m, g)$          & a resource demand vector of a transcoding profile $n$, where: \\ 
                             & $m$ is memory in GB, \\
                             & \[
                                g = \begin{cases}
                                        1, & \text{if GPU should be allocated} \\
                                        0, & \text{otherwise}
                                    \end{cases}
                                \] \\
  $b_{n}$                    & the average outbound bandwidth (GB/sec) of a transcoding profile $n$ \\ 
  $b^{s}_{t}$                & the average inbound bandwidth (GB/sec) that spectator $s$ can contain at time $t$ \\
  $c(\vec{n})$               & the cost (per second) of hosting a transcoding profile $n$ using FaaS  \\
  $o(b_n)$                 & the cost (per GB) of the outbound traffic produced by a transcoding profile $n$ \\ 
  $r^{s}_{t}(QoE)$           & the revenue generated by spectator $s$ at time $t$ for $QoE$ level (see Eqs.~\ref{eq:revenue-constant},~\ref{eq:revenue-linear},~\ref{eq:revenue-logistic}) \\
  $QoE^{s}_{\boldsymbol{K} \rightarrow \boldsymbol{N}}$                      & QoE of a spectator $s$ when consuming transcoding profiles allocation $\boldsymbol{K} \rightarrow \boldsymbol{N}$,~$\forall k \in \boldsymbol{K}$,~$\forall n \in \boldsymbol{N}$ \\ \\

  \hline \hline
  \emph{Decision Variables} & \\
  \hline
  
  $y^{s}_{k \rightarrow n}$ & spectator consumption assignments: \[
                       y^{s}_{k \rightarrow n} =
                             \begin{cases}
                                1, & \text{if spectator} ~s~ \text{consumes profile}  ~n~ \text{for player}~k \\
                                0, & \text{otherwise}
                             \end{cases}
                    \] \\ 
  $f^{s}_{k->n}$  &  $\in (0,1)$:  a fraction of the nominal bandwidth consumption $b_{n}$ of $s$ reduced to match capacity  \\

  \hline \hline
  \emph{Auxiliary Variables} & \\
  \hline
  
  $x_{n}^{k}$ & transcoding profile active status: 
\begin{equation*}
    x_{n}^{k} = \begin{cases}
        1, & \text{if transcoding profile}~n~ \text{is produced for player}~k \\
        0, & \text{otherwise}
    \end{cases}
\end{equation*} \\ 
  \hline \hline
  \end{tabular}
\end{table*}

\noindent \textbf{Given} a set of transcoding profiles $N$, a set of players $K$, and a set of spectators $S_{t_0}$ at time $t_{0}$, \\ \\
\textbf{Determine} the transcoding profiles $x_{n}^{k}$ that should be produced and assign which of those produced profiles each spectator should consume $y^{s}_{k \rightarrow n}$, so as to  \\ \\
\textbf{Maximize} an estimated total profit (ETP) for the immersive gaming service provider:\\

Based on all of the above, the expected total profit is given by:



\begin{equation}
  \begin{split}
    \label{eq:ETP}
    ETP =    &\sum_{s \in \bm{S_{t}}}  E\Bigg(r^s_{t \rightarrow |T|} \bigg( QoE^{s}_t(y^{s}_{\bm{K} \rightarrow \bm{N}} ) \bigg) \Bigg) \\
    - &\sum_{k \in \bm{K}} \sum_{n \in \bm{N}}  \bigg( c(\vec{n}) \cdot \Bigl\lceil \frac{\sum_{s \in \bm{S_{t}}}  y^s_{k \rightarrow n}}{|S|} \Bigr\rceil \bigg)\\
    - & o \cdot \sum_{n \in \bm{N}}\bigg[  b_n \cdot \sum_{k \in \bm{K}}\sum_{s \in \bm{S_{t}}} \bigl( y^s_{k \rightarrow n} - f^{s}_{k \rightarrow n} \bigr)  \bigg] 
  \end{split}
\end{equation}


\textbf{Subject to constraints}:
\begin{equation}
   \sum_{n \in \bm{N}} y^s_{k \rightarrow n} =  1,  ~\forall{s \in \bm{S_t}},~\forall{k \in \bm{K}}, ~\forall{t \in \bm{T}}
 \label{eq:contsraint-correctness-single-allocation-per-spectator}
\end{equation}

\begin{equation}
  \begin{split}
  \label{eq:constraint-feasibility-bandwidth}
    \sum_{k \in \bm{K}} \sum_{n \in \bm{N}} b_n \cdot \bigl( y^s_{k \rightarrow n} - f^{s}_{k \rightarrow n} \bigr) \leq b^s_t,& \\
    \forall{s \in \bm{S_t}}, ~\forall{t \in \bm{T}}, & f^{s}_{k->n} \in (0,1)
  \end{split}    
\end{equation}

\begin{equation}
    y_{k \rightarrow n}^s \geq f_{k \rightarrow n}^{s}, ~\forall{s \in \bm{S_t}},~\forall{k \in \bm{K}}, ~\forall{n \in N}, ~\forall{t \in \bm{T}}
   \label{eq:feasibility-bandwidth-nonnegative}
\end{equation}

\ignore{
\begin{equation}
    x^k_n = sign \left(\sum_{s \in \bm{S_t}} y^s_{k \rightarrow n}\right),~\forall{k \in \bm{K}},~\forall{n \in \bm{N}},~\forall{t \in \bm{T}} 
    \label{eq:produced-consumed}
\end{equation}
}

\ignore{
\begin{equation}
    y^{s}_{k->n} \leq x^{k}_{n}, \forall s \in \bm{S_t}, k \in \bm{K}, n \in \bm{N}
    
\end{equation}
}

Eq.~\ref{eq:ETP} gives the expected total profit (ETP) of the provider as the difference between the expected revenue and the costs. It consists of three terms: The \textbf{first term}, deriving from Eq.~\ref{eq:expected_revenue}, sums the expected revenue for all spectators, which is a function of their QoE, which, in turn, depends on the profiles each of them is assigned to consume. The \textbf{second term} represents transcoding costs, summed for all players and profiles. For each player/profile combination, the ceiling function returns \(1\) if at least one spectator consumes that profile (and hence it is actually in production), and \(0\) if none do. The \textbf{third term} calculates traffic costs, summed for all spectators, players and profiles. Each profile \(n\) has an average bandwidth requirement of \({b_n}\), which is the maximum consumed by a spectator \(s\) who is receiving that profile from player \(k\) (i.e. \(y^s_{k \rightarrow n}=1\) ). However, some of these spectators may be receiving \(n\) at a lower framerate and thus consume less bandwidth; this reduction is expressed by \(f^s_{k \rightarrow n}\).

Constraint~\ref{eq:contsraint-correctness-single-allocation-per-spectator} ensures that each spectator is allocated exactly one transcoding profile per each player.  Constraint~\ref{eq:constraint-feasibility-bandwidth} makes sure that the total effective bandwidth consumed by any spectator at any given time instance does not exceed the maximum bandwidth that this spectator can contain. QoE modeled by Eq.~\ref{eq:qoe} implicitly corrects the frame rate to match constraints of a spectator on inbound bandwidth at time $t$. In the problem formulation we explicitly model this via bandwidth adaptation coefficients $f^s_{k \rightarrow n} \in (0,1)$ . 
Finally, Constraint~\ref{eq:feasibility-bandwidth-nonnegative} prevents negative outbound traffic allocation.

\begin{algorithm}[ht]
\SetAlgoLined

\KwIn{Transcoding Profiles $\bm{N}$, Players $\bm{K}$}
\KwOut{\\
	\hspace*{0.8cm} (1) Optimized Deployment of Transcoding Profiles $\forall t \in \bm{T}$\\
    \hspace*{0.8cm} (2) Optimized Allocation of Transcoding Profiles to spectators $\forall t \in \bm{T}$  
    }

QoE model $\leftarrow$ Eq.~\ref{eq:qoe}\;
Costs model $\leftarrow$ pricing plan for GPU and CPU FaaS and outbound bandwidth\;
Revenue model $\leftarrow$ e.g. Eq.~\ref{eq:revenue-constant}\; 

\For{$t\gets t_0$ \KwTo $|T|$}{
	 $Metrics \leftarrow$ collect metrics from the active spectators $\bm{S_t}$\;
	 Infer $b^s_t$ and compute power $\forall s \in \bm{S_t}$\; 
	 Estimate $QoE^{s}_{\bm{K} \rightarrow \bm{N}}$, $\forall{s} \in \bm{S_t}$ and $\forall \bm{K} \rightarrow \bm{N}$\;
	 Estimate $q_t^s$ $\forall s \in \bm{S_t}$\;
	 $y^{s}_{k \rightarrow n},f^{s}_{k \rightarrow n} \leftarrow$ solve the optimization, considering Eq.~\ref{eq:ETP},
	 Eq.~\ref{eq:contsraint-correctness-single-allocation-per-spectator},
	 Eq.~\ref{eq:constraint-feasibility-bandwidth}, Eq.~\ref{eq:feasibility-bandwidth-nonnegative}\; 
	 Determine which transcoding profiles to activate for each player: 
	 \[  x_{n}^{k} =
                             \begin{cases}
                                1, & \text{if} ~\exists~ y_{k \rightarrow n}^s = 1 \\
                                0, & \text{otherwise}
                            \end{cases}\] \\
     Activate transcoders (via FaaS), as required\;
     Inform spectators about the new profile allocation\;
}

\caption{The Overall Smart network-centric optimization Algorithmic Framework}
\label{alg:frame}

\end{algorithm}

Algorithm~\ref{alg:frame} depicts how we solve the provider problem in the on-line setting. Since in this setting the future is not known, we solve the optimization problem  at every time step, estimating the revenue that will be accumulated if all currently active spectators will remain in the stream.
In the next time step we correct the estimation and again solve the optimization problem to deploy the transcoding profiles and allocate them to the spectators. Since the network conditions (as well as availability of the compute resources) might change from one time window to another for spectator $s$, the transcoding profile allocation for $s$ can also change. As we use FaaS, there are no additional costs associated with releasing serverless transcoders and starting new ones. Since in practical settings the optimization problem is relatively small it can be solved exactly either using linear solvers like CPLEX or even through brute force. 

Obviously, the proposed algorithm is suboptimal, because it is based on estimating the quitting probabilities of spectators based on an estimated QoE and does not make long term decisions. Estimating QoE can be tackled in a number of ways. In~\cite{Athanasoulis2020mmedia} this problem is approached using reinforcement learning, reducing its complexity. 

Achieving cost efficiency depends on an accurate modelling of the costs and revenues. The former depends on the available cloud and 5G MEC commercial offerings for FaaS. The latter depends on the spectators' behavior. The modeling approach of Subsections~\ref{subsec:spectator-behavior}--\ref{subsec:costs} is relatively simple and generic, developed with the use-case of immersive 3D media live streaming in mind. Naturally, each use-case will have its own peculiarities, which will need to be modelled accurately and possibly fine-tuned using real data. 

In this paper, our focus is on demonstrating that even for the relatively simple model, the serverless computing paradigm might result in significant benefits to the provider.

\section{Experiments and Results}
\label{sec:experiments}
We performed a series of experiments to validate the proposed optimization approach and quantify its benefits in different scenarios and conditions. 
Our experiments consider the aforementioned Augmented VR game use-case, in which the spectators must receive two 3D video streams, one for each of the two players.

\subsection{Experimental setup}

To develop and test the application functionality, all components of the service, as described in Section \ref{sec:serverless} and Fig. \ref{fig:service_compoenents}, were implemented, deployed and tested in the infrastructure provided by the 5G-MEDIA project\footnote{http://www.5gmedia.eu/} that offered Kubernetes NFVI with worker nodes equipped with NVIDIA GTX Geforce 1650 GPUs and Open Sorce MANO (OSM) R5.05 with FaaS VIM plugin installed\footnote{https://github.com/5g-media/faas-vim-plugin}. Players, spectators, and  the control plane, have been deployed locally on PCs, while the broker, transcoders and replay application components have been deployed as FaaS VNFs via OSM/FaaS Plugin and orchestrated by the control plane in an event-driven manner using our Serverless Orchestration mechanism with OSM being a unified entry point. 

However, the infrastructure we had access to has been relatively small and imposed hard limits on both the number of spectators and the number of concurrent transcoders that can use GPUs.
Hence, after initial tests on the actual infrastructure, a more extensive study of cost optimization was conducted using simulation. 


\ignore{
The experiments were performed using two versions of the CNO and a number of simulated spectators, with the overall algorithmic framework defined in Algorithm~\ref{alg:frame}.
}

\subsubsection{Simulated spectators}

The Simulated spectators adhere to joining and quitting behavior described in Subsection~\ref{Joining} and Subsection~\ref{Quitting}. The experiments feature a diverse set of spectators, varying in connection bandwidth and processing capabilities, to reflect a mixture of real-life user profiles. 
For each spectator a set of metrics is collected every 10 seconds, reporting their bandwidth, processing power, the transcoding profiles they are currently receiving, and the framerate for each. Based on those metrics, each spectator's current QoE is be calculated (from Eq. \ref{eq:qoe}), as well as an estimate of the QoE they would experience if they were to receive different transcoding profiles.

We consider relatively small sessions and assume that GPUs  are available as needed in 5G MEC's NFVI. 

Spectator bandwidth is subject to a small degree of random fluctuation, to simulate changing network conditions. Likewise, the processing power that can be allocated to the video processing is varied to simulate changing workload conditions of the user equipment. Processing power can impose a limit to the maximum frame-rate a spectator can decode. 

\subsubsection{Network Optimization}

The control plane receives metrics from all spectators and, based on them, decides the optimal set of transcoding profiles that must be produced, and which one of them each spectator should consume. 
The algorithm makes decisions on $10$ second time steps corresponding to the $10$ second monitoring intervals. 

We compare two optimization algorithms:
\begin{itemize}
\item \emph{Naive Optimization} greedily optimizes spectator QoE. Based on the QoE modelling described in Subsection~\ref{QoE_model}, it determines which transcoding profile will result in the optimal QoE for each spectator and allocates transcoders to produce this set of profiles, regardless of the production cost.

\item \emph{Smart Network-Centric Optimization} optimizes cost-efficiency. It balances the trade-off between the profit and QoE.  It considers spectator QoE, the quitting probability as a function of QoE, revenue generated by the spectators remaining online, and production and delivery costs, and determines the set of transcoding profiles to be produced to maximize profit (see Eq.~\ref{eq:ETP}, Eq.~\ref{eq:contsraint-correctness-single-allocation-per-spectator}, Eq.~\ref{eq:constraint-feasibility-bandwidth}, Eq.~\ref{eq:feasibility-bandwidth-nonnegative}). In particular, expected revenue (Eq.~\ref{eq:expected_revenue}) is calculated on the assumption that profiles assigned to spectators during the current time step will also persist for future time steps.
\end{itemize}

In what follows, we will refer to these two algorithms simply as \emph{Naive} and \emph{Smart}.

\subsubsection{Transcoder Parameters}

For each of the two players' 3D video streams five transcoding profiles are supported, in addition to the production streams, which are also available for consumption by the spectators and require no transcoding. 

The production stream and all still images profiles, encode textures as JPEG images of quality 30, in various resolutions.
The video profiles encode textures as a HEVC video of fixed resolution, targeting various bit-rates.
Production frame-rate is set to be 25 frames per second.
Table \ref{table:qualities} lists the specifications of transcoding profiles used in our experiments.
\begin{table}[!htbp]
    \captionsetup{justification=centering}
    \caption{Transcoding profiles' specifications}
    \centering
    \resizebox{\columnwidth}{!}{
    \begin{tabular}{c c c c c c c} 
        & & & \multicolumn{2}{c}{Texture} & \multicolumn{2}{c}{Mesh} \\
        Name & Node & Frame size & Resolution & PSNR & Geometry & Blend weights \\
        \hline \hline
        Production & None & 200 KB & 960x540 & 32.02 dB & 10 bits & 6 bits\\
        \hline
        Images Mid & CPU & 170 KB & 864x486 & 28.78 dB & 9 bits & 5 bits\\
        \hline
        Images Low & CPU & 135 KB & 768x432 & 28.02 dB & 8 bits & 4 bits\\
        \hline
        Video Low & GPU & 55 KB & 960x540 & 28.66 dB & 8 bits & 4 bits\\
        \hline
        Video Mid & GPU & 70 KB & 960x540 & 30.00 dB & 9 bits & 5 bits\\
        \hline
        Video High & GPU & 85 KB & 960x540 & 31.59 dB & 10 bits & 6 bits\\
    \end{tabular}
    \label{table:qualities}
    }
\end{table}

Still images profiles produce slightly worse image quality and, naturally, a significantly larger average frame size. They can be transcoded in real time on a CPU node. Since they have no inter-frame decoding dependency, spectators who fail to receive or decode the frames at the production frame-rate can skip frames to always display the most current frame.

On the other hand, video profiles achieve better image quality with much higher compression rates, and require a GPU node for real-time transcoding. In the test implementation, they do not support skipping frames, due to inter-frame compression. Spectators who cannot match the production framerate may start lagging behind, so the optimization algorithm running in the control plane will never assign a video profile to such a spectator.

\subsubsection{Experiment timeline}

Each experiment considers a streaming session of ten minutes. 
The stream starts with a default number of spectators being $10$. 
New spectators join the stream according to the Poisson process (See Subsection~\ref{Joining}) and leave the stream according to their probability of quitting (see Subsection~\ref{Quitting}).

\begin{figure*}[!htbp]
  \centering
  \includegraphics[width=0.85\textwidth]{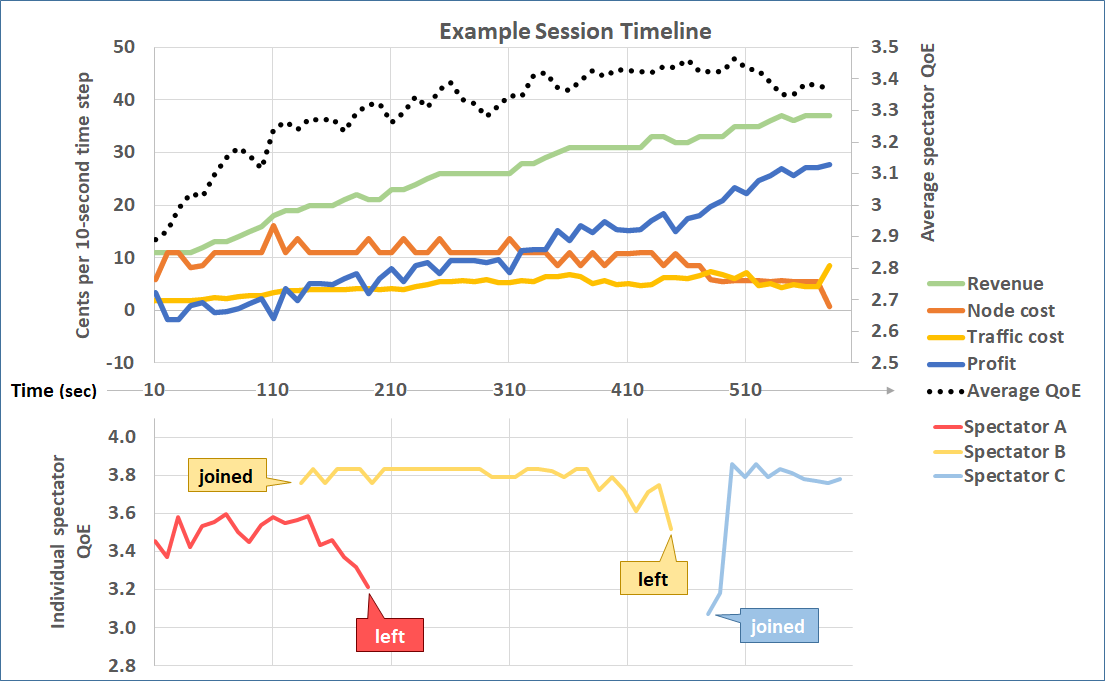}
  \caption{A sample timeline of an experiment. Above the horizontal axis:  revenue, node cost, traffic cost, profit (left vertical axis) and mean QoE (right vertical axis). Below the horizontal axis: Individual QoE progress for three example spectators.}\label{fig:timeline}
\end{figure*}

Figure \ref{fig:timeline} shows the timeline of a sample experiment featuring the Smart optimization in the control plane.
Cost-related quantities (revenue, cost, and profit) correspond to the upper left vertical axis and are shown in a per-time-step basis. Initially revenue is low, due to a low number of initial spectators, resulting in low profit, at times dipping into the negative, for the first couple of minutes. As the session progresses, more spectators join, gradually increasing revenue and profit. 

Node costs are significant in the beginning of the session, becoming less so as more spectators join and revenue increases. They remain more or less constant, while random changes in spectator bandwidth or processing power occasionally result in the production of an extra set of GPU profiles, when it is deemed profitable. Towards the end of the session node costs dip lower, as the expected future revenue of individual spectators diminishes, capped by the session duration. The higher cost of video profiles is partially offset by their lower bitrate, which result in lower traffic costs. The latter naturally increase as more spectators join, but at a much lower rate, reaching a plateau after about four minutes, when the active spectator population can justify the production of more video profiles.

Average QoE follows an upward trend. As more spectators join, the increased revenue can support the production of more transcoding profiles, able to satisfy a more diverse population.

The part of the graph below the horizontal axis shows the QoE progress of three sample spectators. Although the behavior of individual spectators has very small impact to the total revenue and profit in a session of about 20 active spectators, these examples can provide some intuition about the progress of a session.
Spectator A has joined from the start; she has limited bandwidth with significant fluctuations. Sometimes she is unable to receive the better quality profiles, and her QoE drops as a result. 
Finally, when it drops too low, she decides to leave. Her departure can be seen marking a small decrease in the revenue and profit of the next couple of time steps.
Spectator B joins in the middle of the stream and experiences only very minor fluctuations. After some time he leaves, perhaps for non-QoE-related reasons, as his QoE is not so low.
When spectator C joins, she is experiencing a quite low QoE.
However, the optimization algorithm quickly assigns an appropriate transcoding profile that maximizes her QoE, and so she remains active until the end of the stream.

Abrupt changes in individual spectators' QoE are not always reflected in the average QoE or the profit, meaning that the optimizer made a decision to lower some spectators' QoE and raise that of the others, aiming for maximum expected profit. 

\subsubsection{Experiment variables} \label{Variables}

In order to obtain general results and compare the Smart and Naive optimization adaptability under different client conditions, experiments were performed using a range of values for different experiment variables. 
Each set of experiments measured the impact of changing one variable while keeping the others constant at their default values. Experiment variables included:

\renewcommand{\labelitemii}{$\bullet$}

\begin{itemize}
    \item Spectator arrival rate, following the Poisson distribution. Default value of 0.5 new spectators per time step.
    \item Revenue generated by each spectator. Default value of 0.2 cents per time step.
    \item Numbers of GPUs available. Six of the transcoding profiles require a GPU to perform in real time. A limitation on GPUs that can be used concurrently (implies that some profiles may not be produced concurrently). The default number is 6, i.e. no limitation on the concurrent use of GPUs.
    \item GPU costs. As mentioned before, there is currently no commercial option to rent GPU processing for FaaS. Based on calculations derived from current CPU and GPU pricing for VMs, and considering the implementation obstacles in GPU sharing, we estimate a default value of 10 times that of an otherwise equivalent CPU node, and test for factors between x5 and x20, which seem reasonable, considering the analysis in~\ref{pricing:sec}.
    \item Spectator population. We identify 5 broad types of spectators:
    \begin{itemize}
        \item Mobile devices on Wi-Fi.
        \item Mobile devices on 4G data.
        \item Standard PC on basic DSL connection.
        \item Standard PC on faster connections (e.g. VDSL)
        \item High-end PC on a fiber optic connection.
    \end{itemize}
    In the preliminary experiments, average decoding timings for all transcoding profiles were measured for each device type, and these are used to calculate a maximum frame rate from a hardware perspective. In addition, each connection type is associated with a bandwidth typical for it, which provides a frame rate cap from a connection perspective. 
    The default population consists of a balanced mix of the above spectator types, while we also conduct experiments where specific types of spectators are dominant.
    \item Quitting behavior: As mentioned in Subsection~\ref{Quitting}, quitting probability is a function of QoE dissatisfaction and non-QoE-related causes, such as how interesting a specific session is. As the default, derived from \cite{chen2015qoe}, we assume a 20\% probability to quit before the session's end at maximum QoE. We conduct experiments for relatively boring (50\% to quit) or interesting sessions (10\% to quit), and also for more demanding spectators, in which case QoE dissatisfaction weighs more.
\end{itemize}

\subsection{Results} \label{Results}

\begin{figure*}[!h]
  \centering
  \includegraphics[width=0.85\textwidth]{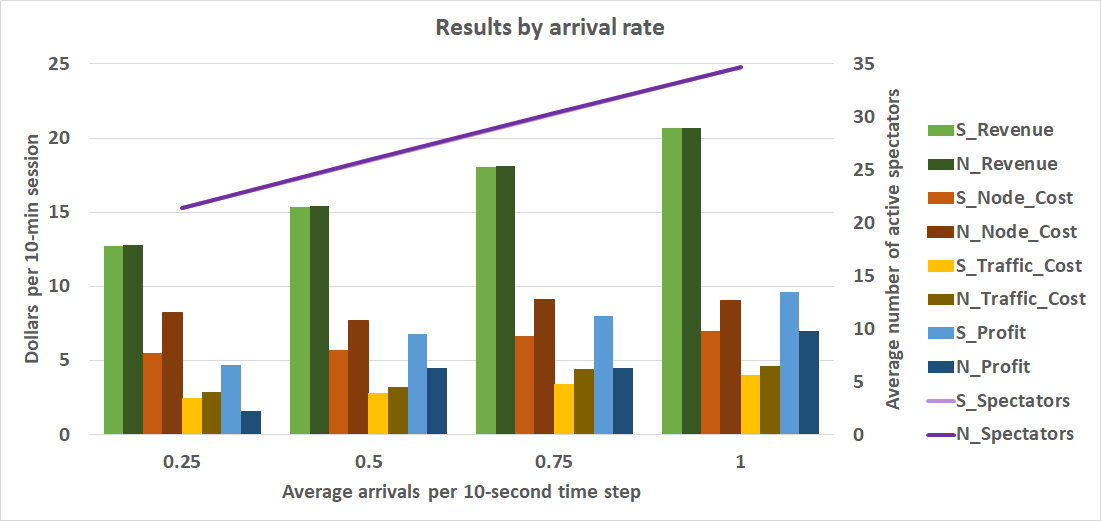}
  \caption{Total session revenue, costs and profit (on the left vertical axis) and average spectator number (on the right vertical axis) for different rates of new spectator arrival.}\label{fig:arrival_rate}
\end{figure*}

In the following graphs we present a comparison between Smart and Naive optimizations. 
Each graph displays a number of experiments differing at one experiment variable (see Subsection~\ref{Variables}), shown on the horizontal axis while keeping the others constant.
The graphs show aggregate measurements of the entire ten-minute sessions, averaged across several replications of the same experiment.

Quantities denoted with an \emph{S} refer to Smart and are shown in a lighter shade, while those denoted with an \emph{N} refer to the Naive. 
Across all graphs, money-related quantities (revenue, cost, and profit) are shown as bars and correspond to the left vertical axis. 
An additional quantity, relevant to each graph, is shown as a line corresponding to the right vertical axis. Such  additional values may include:

\begin{itemize}
    \item Spectators: The average number of spectators active during the session. This directly impacts revenue.
    \item dQoE: the average difference between spectators' actual QoE and the maximum QoE they could possibly achieve, given their network connection and hardware. This directly impacts the quitting probability, which indirectly affects revenue. dQoE is shown in an inverted vertical axis.
    \item Total QoE: The sum of spectators' QoE, averaged across all time steps, which can be perceived as a measurement of a total quality of experience.
\end{itemize}

\begin{figure*}[!h]
  \centering
  \includegraphics[width=0.85\textwidth]{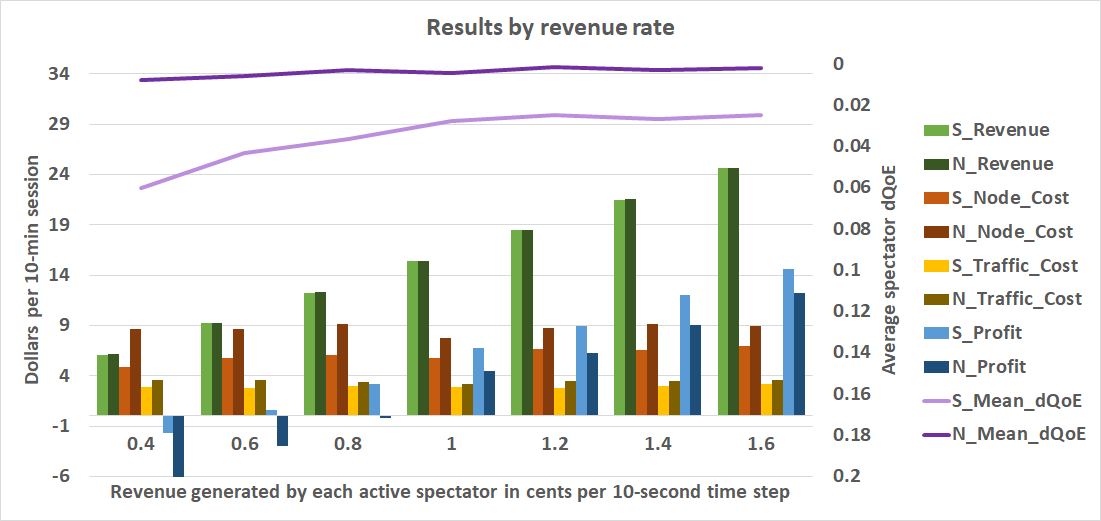}
  \caption{Total session revenue, costs and profit (on the left vertical axis) and average difference from the maximum possible QoE (on the right vertical axis) for the different (constant) rates of revenue per active spectator and per time step.}\label{fig:revenue_rate}
\end{figure*}

\subsubsection{Arrival rate}

Fig.~\ref{fig:arrival_rate} compares performance of Naive and Smart under different spectator arrival rates. 
As the rate increases, so does the average number of spectators staying in the session and the revenue they generate. Traffic costs also naturally increase, as there are more spectators downloading the streams.
With more spectators, Smart's spending on transcoding nodes also increases slightly, as the higher costs of producing more profiles is offset by a larger number of spectators who will benefit from them. It can be noted that, as expected, the Smart's advantage is more pronounced when less spectators are active. 

\begin{figure*}[!h]
  \centering
  \includegraphics[width=0.85\textwidth]{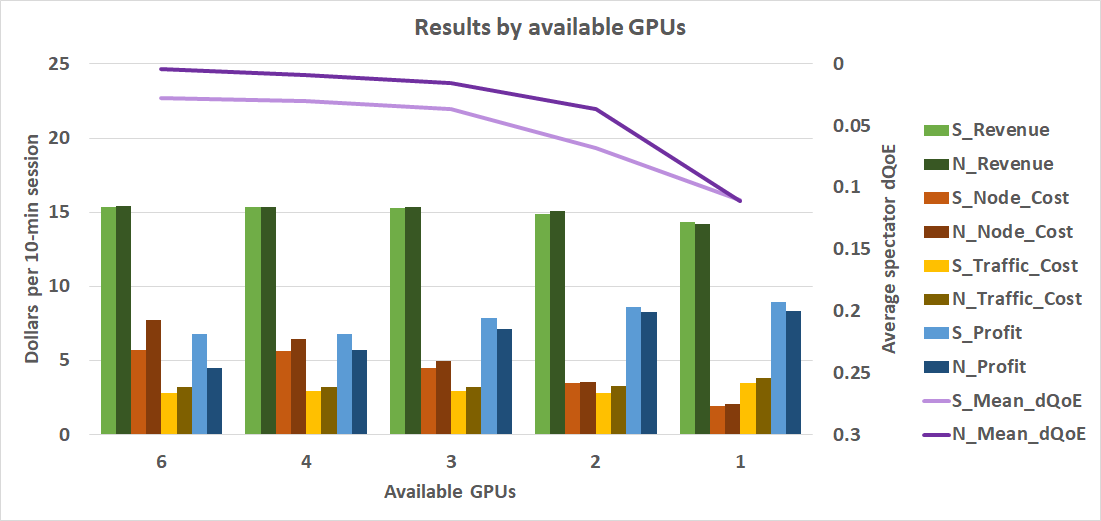}
  \caption{Total session revenue, costs and profit (on the left vertical axis) and average difference from the maximum possible QoE (on the right vertical axis) when different numbers of GPUs are available for the transcoding of streams to video in real time.}\label{fig:GPUs}
\end{figure*}

\subsubsection{Revenue rate}

Fig.~\ref{fig:revenue_rate} presents the experimental results for different revenue rates around the default of 1 cent per 10-second time step. This set of experiments follows the default arrival rate of 0.5, meaning that revenue is generated by an average of about 25 spectators per session.

Naturally, as the revenue generated by each active spectator increases, so too does the overall revenue. The costs  also increase, as it is becomes more profitable to keep spectators satisfied. 
This is also reflected on the decreasing dQoE, shown on the right vertical axis. 
At lower revenue rates Smart makes a greater difference in profit. 
A similar behavior of Smart is observed with the increasing arrival rate as shown in Fig.~\ref{fig:arrival_rate}. 

\subsubsection{Available GPUs}

This series of experiments, shown in Fig.~\ref{fig:GPUs}, considers the case where production GPUs usage is limited, capping the number of video profiles that can be transcoded simultaneously. As the number of available GPUs drops, so do the options and versatility of Smart, limiting its benefit.

\subsubsection{Revenue model}


\begin{figure*}[!htbp]
  \centering
  \includegraphics[width=0.85\textwidth]{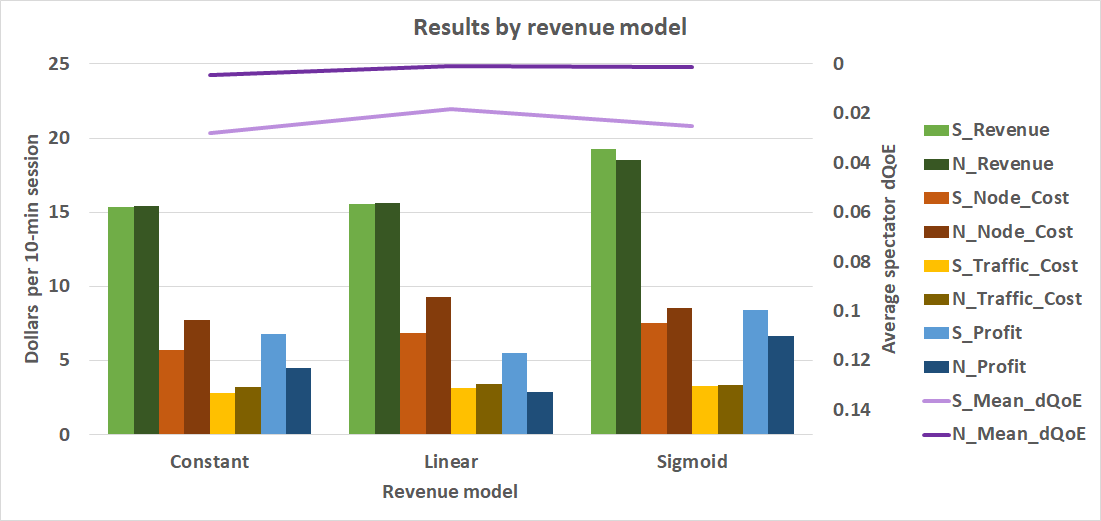}
  \caption{Total session revenue, costs and profit (on the left vertical axis) and average difference from the maximum possible QoE (on the right vertical axis) is we assume a constant revenue rate (regardless of QoE) or a revenue rate that is a linear or sigmoid function of QoE, rewarding better QoE with higher revenue.}\label{fig:revenue_model}
\end{figure*}

Fig. \ref{fig:revenue_model} compares system behavior with different revenue models regarding spectator QoE.
With the constant model spectators deliver a set revenue so long as they remain online, while with the linear and sigmoid models spectators with good QoE generate more revenue than ones with bad QoE. 
This makes good QoE more important, as it impacts revenue both directly and indirectly (by affecting quitting probability). 
With the linear and sigmoid models revenue models the advantage of Smart over Naive is less pronounced since it tends to maximize QoE similarly to Naive.

\subsubsection{GPU node cost}

\begin{figure*}[!htbp]
  \centering
  \includegraphics[width=0.85\textwidth]{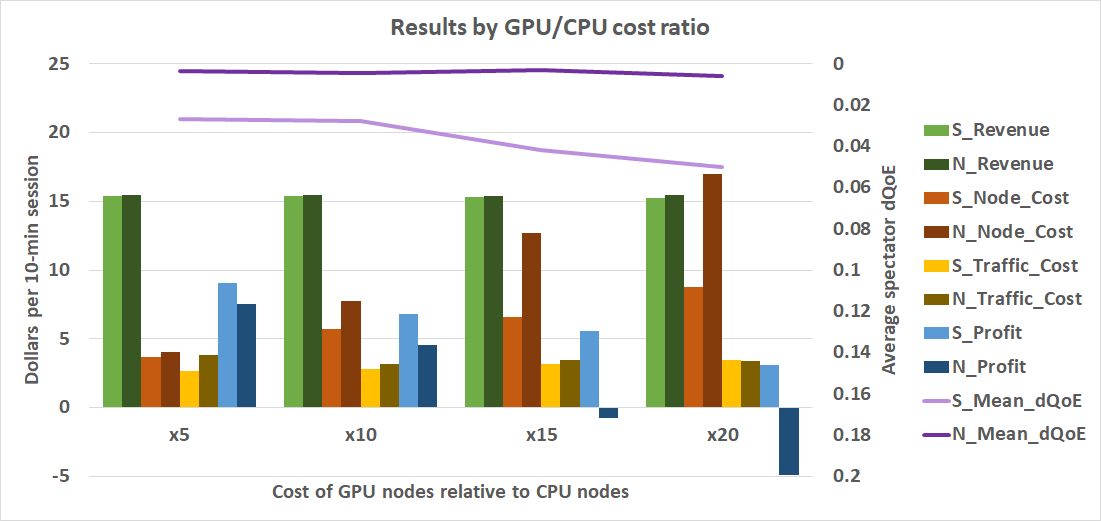}
  \caption{Total session revenue, costs and profit (on the left vertical axis) and average difference from the maximum possible QoE (on the right vertical axis) if we assume different ratios in the prices of GPU nodes to CPU nodes. Ratios of x5 - x20 seem reasonable, based on current VM rental prices.}\label{fig:GPU_cost}
\end{figure*}

As mentioned before, to the best of our knowledge there is currently no commercial option to rent GPU nodes for FaaS processing.
However, it is entirely possible that such options will be available in the near future, especially if demand for it rises. GPU processing will certainly cost more than CPU processing. 
This set of experiments examines the impact of the price ration between GPU and CPU nodes. 
As seen on Fig.~\ref{fig:GPU_cost}, as GPU processing becomes more expensive, Smart becomes more frugal with GPU-dependent profiles, letting spectator QoE drop away from the optimal. Hence, it can keep running costs manageable and generate a profit even when GPU utilization is priced high, with a small decrease in QoE. It can be noted that although at low GPU pricing optimization offers a relatively small benefit, this becomes much more emphasized when GPU usage is more expensive. Also note how Smart's traffic costs get higher as GPU cost increases and still-image CPU profiles, which have a lower compression rate and thus higher bitrate, are preferred.

\subsubsection{Spectator population}

\begin{figure*}[!htbp]
  \centering
  \includegraphics[width=0.85\textwidth]{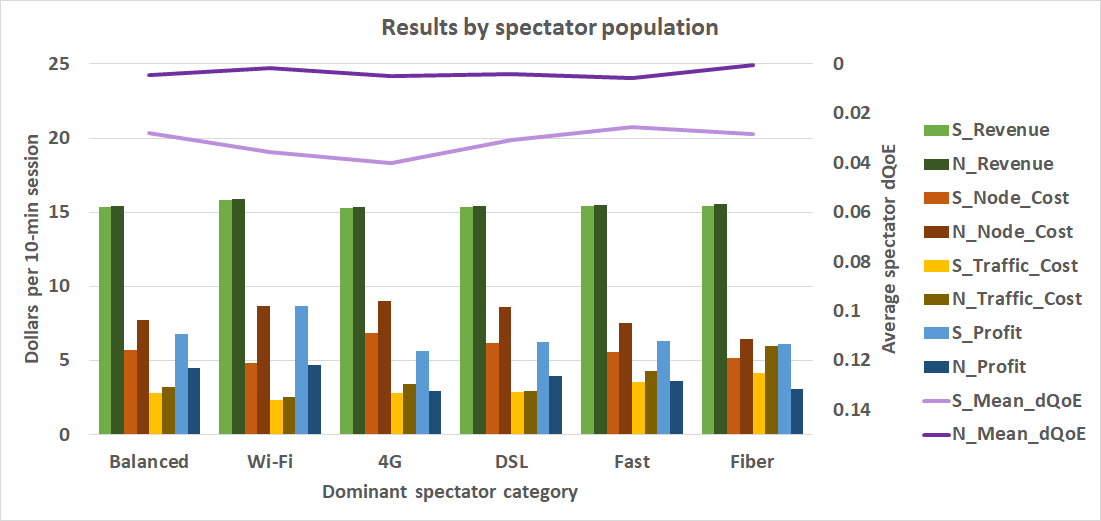}
  \caption{Total session revenue, costs and profit (on the left vertical axis) and total aggregate QoE (on the right vertical axis) for spectator populations where different types are more frequent.}\label{fig:spectators}
\end{figure*}

Fig. \ref{fig:spectators} regards experiments with different spectator populations. 
Although all experiments contain all types of spectators, in this set we examine the impact of having different dominant spectator types.
Smart holds a steady advantage across all cases. The two right-most sets of bars, corresponding to a greater percentage of faster connections, shows a marked decrease in node costs, offset by an increase in traffic costs, as many of those spectators can consume the high-bitrate production stream, obviating the need for (and cost of) transcoding.

\subsubsection{Quitting probability}

\begin{figure*}[!htbp]
  \centering
  \includegraphics[width=0.85\textwidth]{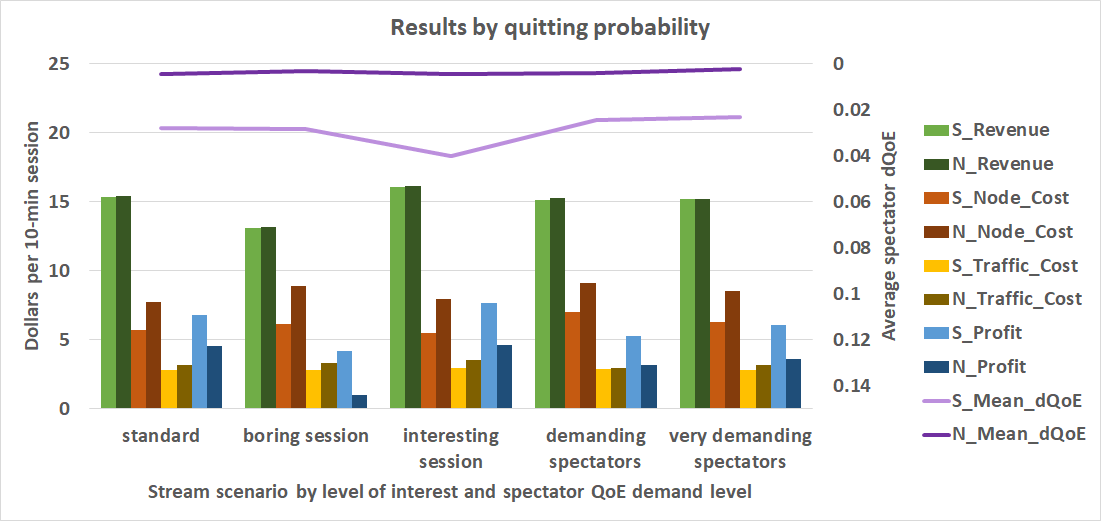}
  \caption{Total session revenue, costs and profit (on the left vertical axis) and average difference from the maximum possible QoE (on the right vertical axis) when different factors affect spectator quitting probability, including how interesting or boring the session is and how demanding the spectators.}\label{fig:quitting}
\end{figure*}

This set of experiments considers how quitting probability impacts Smart's decisions. Fig.~\ref{fig:quitting} shows measurements for standard boring and interesting sessions, in which quitting probability is respectively higher or lower; and also how more demanding spectators (in which case QoE dissatisfaction weighs more in their probability of quitting) affect the process. 
In a boring session, Smart increases spending in an effort to keep spectators from quitting via providing a better QoE.
On the contrary, in the interesting session Smart reduces costs and allows QoE to drop, as spectators are less likely to leave anyway. With more demanding spectators, Smart targets higher QoE to keep them engaged, resulting in higher transcoding node costs.

\subsubsection{Summary of the results}

Having conducted experiments with parameters spanning numerous different assumptions and cases, some overall conclusions may be reached.
Smart QoE-cost optimization can reduce transcoding costs by up to 60\% and traffic costs by about 20\%, while keeping revenue and QoE very close to the optimum. Optimization's benefits are especially pronounced in cases with few spectators, low revenue and high GPU costs. 
When GPUs become available for FaaS, possibly in the near future, they can be expected to start at higher prices, gradually dropping as their use becomes more common. Live streaming platforms, especially those dealing in emerging media, will likely need to start with a small spectator base before gaining momentum and scaling. Hence, smart QoE-cost optimization will be indispensable in the live media streaming landscape of the near future.

\section{Conclusion}
\label{sec:conclusion}
5G networks will disrupt the way media intensive services are being developed and operated by unlocking a plethora of opportunities to both service developer and service operator. In this work, we studied one such capability that integrates modern serverless technology with real-time adaptive media streaming in 5G MEC. To the best of our knowledge, this is the first work that does this.

Apart from the conceptual, architectural and technical contributions, our work further examined the potential of this option in terms of network-centric service cost optimization. Our findings indicate that for small user populations and finite duration sessions, serverless adaptive streaming can offer reduced operating expenditure (OPEX) while preserving the service's QoE.

Also, through our extensive modelling and analysis we concluded that naively applying serverless will not necessarily offer these gains.
We hope that our work will inspire further research and development towards adapting services not originally suited for lighter-weight virtualization to serverless architectures, and unify them with the advanced capabilities that 5G networks offer to capitalize on its advantages in novel ways.
Finally, taking into account the recent introduction and the emerging availability of GPUs specifically designed for data centers our work can be extended to accommodate these developments.
Specifically for media services, GPU slicing can allow for even finer-grained cost optimization, something that was not possible before.

One interesting future work direction is to explore more sophisticated placement schemes for transcoders and other components, in which they can be spread across the full compute spectrum across cloud and edge to leverage differentiated pricing for compute, storage and network resources to meet demanding KPIs at lower price points.

\section*{Conflict of interest}
The authors declare that they have no conflict of interest.

\bibliographystyle{spmpsci}      
\bibliography{main}

\end{document}